\documentclass[12pt,english]{article}
\usepackage{times}
\usepackage[T1]{fontenc}
\usepackage{geometry}
\usepackage{rotating}
\usepackage{graphicx}
\usepackage{setspace}
\usepackage{amssymb}

\makeatletter

\providecommand{\tabularnewline}{\\}

\usepackage{bm}

\usepackage{babel}
\makeatother
\begin{document}

\section*{An Opportunistic Source Synthesis Method for Smart Electromagnetic
Environments}

\noindent \vfill

\noindent P. Da Ru,$^{(1)(2)}$ A. Benoni,$^{(1)(2)}$ \emph{Member},
\emph{IEEE}, M. Salucci,$^{(1)(2)}$ \emph{Senior Member}, \emph{IEEE},
P. Rocca,$^{(1)(2)(3)}$ \emph{Fellow, IEEE}, and A. Massa,$^{(1)(2)(4)(5)(6)}$
\emph{Fellow, IEEE}

\noindent \vfill

\noindent {\scriptsize $^{(1)}$} \emph{\scriptsize ELEDIA Research
Center} {\scriptsize (}\emph{\scriptsize ELEDIA}{\scriptsize @}\emph{\scriptsize UniTN}
{\scriptsize - University of Trento)}{\scriptsize \par}

\noindent {\scriptsize DICAM - Department of Civil, Environmental,
and Mechanical Engineering}{\scriptsize \par}

\noindent {\scriptsize Via Mesiano 77, 38123 Trento - Italy}{\scriptsize \par}

\noindent \textit{\emph{\scriptsize E-mail:}} {\scriptsize \{}\emph{\scriptsize pietro.daru,
arianna.benoni, marco.salucci, paolo.rocca, andrea.massa}{\scriptsize \}@}\emph{\scriptsize unitn.it}{\scriptsize \par}

\noindent {\scriptsize Website:} \emph{\scriptsize www.eledia.org/eledia-unitn}{\scriptsize \par}

\noindent {\scriptsize ~}{\scriptsize \par}

\noindent {\scriptsize $^{(2)}$} \emph{\scriptsize CNIT - \char`\"{}University
of Trento\char`\"{} ELEDIA Research Unit }{\scriptsize \par}

\noindent {\scriptsize Via Sommarive 9, 38123 Trento - Italy}{\scriptsize \par}

\noindent {\scriptsize Website:} \emph{\scriptsize www.eledia.org/eledia-unitn}{\scriptsize \par}

\noindent {\scriptsize ~}{\scriptsize \par}

\noindent {\scriptsize $^{(3)}$} \emph{\scriptsize ELEDIA Research
Center} {\scriptsize (}\emph{\scriptsize ELEDIA}{\scriptsize @}\emph{\scriptsize XIDIAN}
{\scriptsize - Xidian University)}{\scriptsize \par}

\noindent {\scriptsize P.O. Box 191, No.2 South Tabai Road, 710071
Xi'an, Shaanxi Province - China }{\scriptsize \par}

\noindent {\scriptsize E-mail:} \emph{\scriptsize paolo.rocca@xidian.edu.cn}{\scriptsize \par}

\noindent {\scriptsize Website:} \emph{\scriptsize www.eledia.org/eledia-xidian}{\scriptsize \par}

\noindent {\scriptsize ~}{\scriptsize \par}

\noindent {\scriptsize $^{(4)}$} \emph{\scriptsize ELEDIA Research
Center} {\scriptsize (}\emph{\scriptsize ELEDIA}{\scriptsize @}\emph{\scriptsize UESTC}
{\scriptsize - UESTC)}{\scriptsize \par}

\noindent {\scriptsize School of Electronic Science and Engineering,
Chengdu 611731 - China}{\scriptsize \par}

\noindent \textit{\emph{\scriptsize E-mail:}} \emph{\scriptsize andrea.massa@uestc.edu.cn}{\scriptsize \par}

\noindent {\scriptsize Website:} \emph{\scriptsize www.eledia.org/eledia}{\scriptsize -}\emph{\scriptsize uestc}{\scriptsize \par}

\noindent {\scriptsize ~}{\scriptsize \par}

\noindent {\scriptsize $^{(5)}$} \emph{\scriptsize ELEDIA Research
Center} {\scriptsize (}\emph{\scriptsize ELEDIA@TSINGHUA} {\scriptsize -
Tsinghua University)}{\scriptsize \par}

\noindent {\scriptsize 30 Shuangqing Rd, 100084 Haidian, Beijing -
China}{\scriptsize \par}

\noindent {\scriptsize E-mail:} \emph{\scriptsize andrea.massa@tsinghua.edu.cn}{\scriptsize \par}

\noindent {\scriptsize Website:} \emph{\scriptsize www.eledia.org/eledia-tsinghua}{\scriptsize \par}

\noindent {\scriptsize ~}{\scriptsize \par}

\noindent {\scriptsize $^{(6)}$} \emph{\scriptsize }{\scriptsize School
of Electrical Engineering}{\scriptsize \par}

\noindent {\scriptsize Tel Aviv University, Tel Aviv 69978 - Israel}{\scriptsize \par}

\noindent \textit{\emph{\scriptsize E-mail:}} \emph{\scriptsize andrea.massa@eng.tau.ac.il}{\scriptsize \par}

\noindent {\scriptsize Website:} \emph{\scriptsize https://engineering.tau.ac.il/}{\scriptsize \par}

\noindent \vfill

\noindent \textbf{\emph{This work has been submitted to the IEEE for
possible publication. Copyright may be transferred without notice,
after which this version may no longer be accessible.}}

\noindent \vfill

\newpage
\section*{An Opportunistic Source Synthesis Method for Smart Electromagnetic
Environments}

\noindent ~

\noindent ~

\noindent ~

\begin{flushleft}P. Da Ru, A. Benoni, M. Salucci, P. Rocca, and A.
Massa\end{flushleft}

\vfill

\begin{abstract}
\noindent In the framework of the {}``Smart ElectroMagnetic Environment''
(\emph{SEME}), an innovative strategy leveraging \emph{Equivalence
Source} concepts is introduced for enhancing the performance of large-scale
outdoor wireless communication systems. The proposed Opportunistic
Sources Synthesis (\emph{OSS}) approach is aimed at unconventionally
synthesizing the primary source (i.e., the base transceiver station
(\emph{BTS}) antenna array), so that the complex scattering phenomena
induced in the surrounding scatterers are profitably exploited to
enhance the received power within user-defined regions of interest
(\emph{RoI}s). To yield a computationally feasible synthesis process,
an innovative \emph{{}``Embedded-plus-Environment Patterns''} (\emph{EPEP}s)
method is introduced. A set of representative numerical examples,
concerned with realistic large-scale outdoor scenarios, is presented
to assess the effectiveness and the efficiency of the proposed optimization-driven
approach for a realistic \emph{SEME} implementation.

\noindent \vfill
\end{abstract}
\noindent \textbf{Key words}: Smart ElectroMagnetic Environment (\emph{SEME}),
Opportunistic Sources Synthesis (\emph{OSS}), outdoor communications,
Ray Tracing (\emph{RT}), Particle Swarm Optimization (\emph{PSO}).

\newpage
\section{Introduction and Motivation}

Nowadays, the limited availability of power and bandwidth as well
as the ever-increasing number of connected end-users and devices are
key challenges to be carefully addressed in designing future wireless
communication systems \cite{Rappaport 2017}-\cite{Geraci 2022}.
Dealing with indoor scenarios, connectivity, reliability, and security
requirements are becoming more and more stringent in both domestic
(e.g., home automation and smart household appliances \cite{Jabbar 2019}\cite{Dong 2021})
and industrial contexts (e.g., Industry 4.0 \cite{Rikalovic 2022}).
Similarly, outdoor systems will have to fulfill the needs of a growing
population of users equipment and to offer seamless connectivity to
stationary and moving targets through disruptive technologies such
as massive multi-user multiple-input and multiple-output (\emph{MU-MIMO})
\cite{Elijah 2022} and their profitable integration with artificial
intelligence (\emph{AI}) \cite{Zardi 2021}. Therefore, future wireless
infrastructures will have to provide the necessary electromagnetic
(\emph{EM}) coverage to yield high data rates, high-reliability, and
low-latency communications, which are mandatory requisites not only
for data-hungry services such as \emph{4K}-streaming, video calling,
and data transfer, but also for emerging paradigms such as tactile
internet \cite{Sachs 2019}\cite{Jebbar 2022}, fog and edge computing
\cite{Hassan 2019}-\cite{McEnroe 2022}, \emph{AI}- and deep learning
(\emph{DL})-based intelligent services \cite{Yu 2021}, and new applications
based on high-precision user localization \cite{Kim 2022}\cite{Torsoli 2023}.
However, traditional solutions adopted in the past by network designers
and planners (e.g., increasing the number and power of transmitting
antennas and/or widening the allocated bandwidth for a specific service)
are no longer feasible due to the stringent regulations on \emph{EM}
compatibility/emission for modern base transceiver stations (\emph{BTS}s),
the congestion of the electromagnetic (\emph{EM}) spectrum, and the
cost for a more pervasive deployment of \emph{BTS}s due to the increase
of the operating frequency \cite{Chiaraviglio 2022}\cite{Massa 2021}. 

\noindent To overcome such a bottleneck, the implementation of the
so-called {}``Smart ElectroMagnetic Environments'' (\emph{SEME}s)
has been recently proposed. Such an approach is aimed at exploiting
in an opportunistic way the propagation environment and the objects
therein, by also adding some smart \emph{EM} entities (\emph{SEE}s),
to enhance the overall performance of a wireless system \cite{Massa 2021}-\cite{Flamini 2022}.
Although the term {}``\emph{SEME}'' encompasses a wide variety of
innovative techniques and technologies, significant efforts have been
recently focused on the design of cost-effective field manipulation
devices (\emph{FMD}s) and their optimal planning in the propagation
environment. More specifically, static-passive electromagnetic skins
(\emph{SP}-\emph{EMS}s) \cite{Oliveri 2023}-\cite{Benoni 2022} and
reconfigurable-passive \emph{EMS}s (\emph{RP}-\emph{EMS}s) \cite{Di Renzo 2019}\cite{Di Renzo 2020}-\cite{Oliveri 2022b}
have been widely studied to leverage their capability of breaking
the traditional Snell's laws \cite{Barbuto 2022} to counteract the
undesired phenomena that negatively impact on the quality-of-service
(\emph{QoS}) (e.g., non-line-of-sight (\emph{NLOS}), shadowing, and
fading \cite{Salucci 2023}\cite{Benoni 2022}).

\noindent In this paper, the implementation of \emph{SEME}s in large-scale
outdoor wireless systems is addressed without adding \emph{SEE}s to
the environment, but directly exploiting the existing propagation
environment in an \emph{opportunistic} way to yield the desired received
power distribution within a user-defined region-of-interest (\emph{RoI}).
More specifically, the proposed approach leverages and properly extends
concepts drawn from the \emph{Inverse Source} theory by assuming that
it is possible to synthesize suitable equivalent currents induced
on the buildings (i.e., the {}``opportunistic sources'') by properly
designing the primary source that illuminates the scenario at hand.
From a methodological viewpoint, the \emph{SEME} is here implemented
by considering the \emph{BTS} antenna perspective. Indeed, the \emph{BTS}
antenna becomes {}``smart'' instead of the {}``environment'' since
it is requested to properly reconfigure the radiated power distribution
for tailoring the scattering phenomena with the scatterers in a profitable
way. While such an idea was preliminarily conceptualized in \cite{Massa 2021},
but it was limited to very simple and small-scale {}``toy examples'',
this is the first time to the best of the authors' knowledge that
the opportunistic source synthesis (\emph{OSS}) paradigm is proposed
as a systematic tool to deploy \emph{SEME}s within real-world large-scale
outdoor scenarios. 

\noindent The main contributions of this work over the existing literature
consist in (\emph{i}) the development of an optimization-driven method
for the coverage improvement in large-scale complex-scattering scenarios
thanks to the opportunistic exploitation of the existing propagation
environment and (\emph{ii}) the introduction of an innovative approach
for reliably predicting in a computationally efficient way the received
power distribution generated by the interaction of the field radiated
by the \emph{BTS} antenna array and the surrounding scatterers. 

\noindent The paper is organized as follows. Section \ref{sec:Mathematical-Formulation}
presents the mathematical formulation of the \emph{OSS} Problem (\emph{OSSP}),
while its solution, based on an optimization approach, is detailed
in Sect. \ref{sub:OSSP-Solution-Methodology}. A set of representative
numerical results is discussed in Sect. \ref{sec:Numerical-Results}
to firstly assess the effectiveness of the \emph{EPEP}-based strategy
and then to prove the capabilities and the potentialities of the developed
\emph{OSS} solution towards the implementation of the \emph{SEME}
paradigm in realistic large-scale outdoor wireless communication scenarios.
Eventually, some conclusions and final remarks are drawn (Sect \ref{sec:Conclusions}).

\section{Problem Formulation (\emph{OSSP})\label{sec:Mathematical-Formulation}}

Let us consider a primary \emph{EM} source modeled as a current density
distribution $\underline{J}_{\Psi}\left(\underline{r}\right)$, defined
on a bounded support $\Psi$ with barycenter $\underline{r}_{\Psi}=\left(x_{\Psi},\, y_{\Psi},\, z_{\Psi}\right)$,
that radiates the \emph{Incident} electric field $\underline{E}_{\Psi}\left(\underline{r}\right)$
at frequency $f$%
\footnote{\noindent Under the hypothesis of time-harmonic \emph{EM} fields,
the scattering phenomena at frequency $f$ can be expressed in terms
of the electric field phasor $\underline{E}\left(\underline{r}\right)=\sum_{\gamma=\left\{ x,y,z\right\} }E_{\gamma}\left(\underline{r}\right)\underline{u}_{\gamma}$,
$\underline{u}_{\gamma}$ being the unit vector along the $\gamma$-th
($\gamma=\left\{ x,y,z\right\} $) Cartesian axis, corresponding to
the time-domain representation $\underline{E}\left(\underline{r},\, t\right)=\Re\left\{ \underline{E}\left(\underline{r}\right)\exp\left(j2\pi ft\right)\right\} $,
where $\Re\left\{ \,.\,\right\} $ is the real part and $j=\sqrt{-1}$
is the imaginary unit.%
}. The arising \emph{EM} interactions between the field radiated by
$\underline{J}_{\Psi}\left(\underline{r}\right)$ and the known arbitrary
arrangement of electrically-large scatterers of a three-dimensional
(\emph{3D}) scenario $\mathcal{D}$ generate the \emph{Total} electric
field distribution \cite{Chen 2018}\begin{equation}
\underline{E}\left(\underline{r}\right)=\underline{E}_{\Psi}\left(\underline{r}\right)+k_{0}^{2}\int_{\mathcal{D}}\underline{J}_{eq}\left(\underline{r}\right)\cdot\underline{\underline{\mathcal{G}}}\left(\left.\underline{r}\right|\underline{r}'\right)d\underline{r}'\label{eq:total-field}\end{equation}
where $k_{0}=2\pi f\sqrt{\varepsilon_{0}\mu_{0}}$, $\varepsilon_{0}=8.85\times10^{-12}$
{[}F/m{]}, $\mu_{0}=4\pi\times10^{-7}$ {[}H/m{]}, $\underline{\underline{\mathcal{G}}}\left(\left.\underline{r}\right|\underline{r}'\right)$
is the dyadic Green's function \cite{Chen 2018}, and \begin{equation}
\underline{J}_{eq}\left(\underline{r}\right)=\tau\left(\underline{r}\right)\underline{E}\left(\underline{r}\right)\label{eq:}\end{equation}
is the equivalent current induced on the scatterers in $\mathcal{D}$.
The \emph{EM} properties of these latters are modeled by the \emph{Object
Function}\begin{equation}
\tau\left(\underline{r}\right)=\left[\varepsilon_{r}\left(\underline{r}\right)-1\right]+j\frac{\sigma\left(\underline{r}\right)}{2\pi f\varepsilon_{0}}\label{eq:}\end{equation}
where $\varepsilon_{r}\left(\underline{r}\right)\geq1$ and $\sigma\left(\underline{r}\right)\geq0$
{[}S/m{]} are the relative permittivity and the conductivity, respectively.
In the generic position $\underline{r}\in\mathcal{D}$, the received
power is equal to \cite{Conciauro 1993}\begin{equation}
P_{rx}\left(\underline{r}\right)=\left[\left|E_{x}\left(\underline{r}\right)\right|^{2}+\left|E_{y}\left(\underline{r}\right)\right|^{2}+\left|E_{z}\left(\underline{r}\right)\right|^{2}\right]\times\frac{\lambda^{2}G_{rx}}{8\pi\eta_{0}}\label{eq:received-power}\end{equation}
where $\lambda$ is the wavelength, $G_{rx}$ is the receiver gain,
and $\eta_{0}=\sqrt{\frac{\mu_{0}}{\varepsilon_{0}}}$.

\noindent Under the above assumptions, the \emph{Opportunistic Sources
Synthesis} (\emph{OSS}) problem can be defined as the synthesis of
the primary source $\underline{J}_{\Psi}\left(\underline{r}\right)$
so that the induced source on the obstacles (i.e., $\underline{J}_{eq}\left(\underline{r}\right)$
$\underline{r}\in\mathcal{D}$) generates a target power distribution
$P_{rx}^{tar}\left(\underline{r}\right)$ in a user-defined \emph{Region
of Interest} (\emph{RoI}) $\Omega\subset\mathcal{D}$ with barycenter
$\underline{r}_{\Omega}=\left(x_{\Omega},y_{\Omega},z_{\Omega}\right)$. 

\noindent In a modern outdoor wireless communications scenario, $\mathcal{D}$
models a typical urban configuration where the \emph{EM} scatterers
are represented by houses and tenements {[}Fig. 1(\emph{a}){]}. Moreover,
the primary \emph{EM} source is a planar \emph{BTS} array of $N$
elements implementing one sector inside a standard hexagonal-cell
network \cite{Baltzis 2011}. Such an array is described by its complex
excitation coefficients\begin{equation}
\underline{w}=\left\{ w_{n}=\alpha_{n}\times\exp\left(j\beta_{n}\right);\, n=1,...,N\right\} \label{eq:}\end{equation}
where $\alpha_{n}$ and $\beta_{n}$ denote the magnitude and the
phase of the excitation of the $n$-th ($n=1,...,N$) radiator within
the aperture $\Psi$, respectively. Moreover, the \emph{BTS} array
is supposed to have the broadside direction mechanically oriented
towards a fixed angular direction $\left(\varphi_{\Psi},\,\theta_{\Psi}\right)$,
where $\varphi_{\Psi}$ is the azimuth angle of the sector ($\varphi_{\Psi}=0$
{[}deg{]} corresponding to the North or, equivalently, to the $y$-axis)
and $\theta_{\Psi}\geq0$ {[}deg{]} is the \emph{BTS} down-tilt angle
{[}Fig. 1(\emph{b}){]} \cite{Yang 2019}. 

\noindent Accordingly, the problem of finding the optimal primary
source, $\underline{J}_{\Psi}^{opt}\left(\underline{r}\right)$, turns
out to be that of deriving the optimal set of \emph{BTS} excitations,
$\underline{w}^{opt}$. Since in practical situations the maximization
of the overall efficiency is generally looked for, the \emph{BTS}
excitations are not tapered and a uniform amplitude distribution is
assumed for the feeding magnitudes. Accordingly, the magnitude of
each $n$-th ($n=1,...,N$) excitation, $\alpha_{n}$, is set to $\xi$
where \begin{equation}
\xi=\sqrt{\frac{\zeta^{\max}}{\left.\zeta\left(\underline{w}\right)\right|_{\underline{w}=\underline{w}_{0}}}},\label{eq:excitation-magnitude}\end{equation}
 $\zeta^{\max}$ being the \emph{BTS} maximum radiated power complying
with local emission masks/regulations and/or system requirements \cite{3G Americas}\cite{4G Americas},
while $\underline{w}_{0}$ is the uniform excitations vector, $\underline{w}_{0}=\left\{ w_{n}=1;\,\,\, n=1,...,N\right\} $,
and $\zeta\left(\underline{w}\right)$ is the radiated power by an
array with excitation vector $\underline{w}$ \cite{Balanis 2005}
\begin{equation}
\begin{array}{r}
\zeta\left(\underline{w}\right)\approx\frac{1}{2\eta_{0}}\int_{0}^{2\pi}\int_{0}^{\pi}\left[\left|E_{\Psi,\theta}^{FF}\left(\left.\theta,\varphi\right|\underline{w}\right)\right|^{2}+\right.\\
\left.+\left|E_{\Psi,\varphi}^{FF}\left(\left.\theta,\varphi\right|\underline{w}\right)\right|^{2}\right]\sin\theta d\theta d\varphi\end{array}\label{eq:Balanis-Radiated-Power}\end{equation}
where $E_{\Psi,\chi}^{FF}\left(\left.\theta,\varphi\right|\underline{w}\right)$
is the $\chi$-th ($\chi=\left\{ \theta,\varphi\right\} $) far-field
(\emph{FF}) component of the \emph{BTS} radiated field along the $\underline{u}_{\theta}$
and $\underline{u}_{\varphi}$ directions, respectively \cite{Ludwig 1973}.
Owing to such hypotheses, the unique degrees-of-freedom (\emph{DoF}s)
of the synthesis problem at hand turn out to be the $N$ phase coefficients\begin{equation}
\underline{\beta}=\left\{ \beta_{n};\, n=1,...,N\right\} .\label{eq:DoFs}\end{equation}
Consequently, the ultimate goal is to synthesize the phase vector
$\underline{\beta}$ to opportunistically exploit the \emph{EM} propagation
phenomena occurring in the surrounding environment (e.g., canyoning,
wave-guiding, scattering, multi-path) for obtaining a distribution
of the received power in a given \emph{RoI}, $P_{rx}\left(\underline{r}\right)$,
$\underline{r}\in\Omega$, that fulfills a user-defined target coverage,
$P_{rx}^{tar}\left(\underline{r}\right)$, $\underline{r}\in\Omega$.
Formally, the synthesis problem at hand can be stated as follows:

\begin{quotation}
\noindent \textbf{\emph{OSS Problem}} \textbf{(}\textbf{\emph{OSSP}}\textbf{)
-} Given an antenna array of $N$ elements operating in a rich-scattering
urban scenario $\mathcal{D}$ with maximum radiated power $\zeta^{\max}$
($\Rightarrow$ $\alpha_{n}=\xi$, $n=1,...,N$) and a target received
power distribution $P_{rx}^{tar}\left(\underline{r}_{m}\right)$ ($\underline{r}_{m}\in\Omega$,
$m=1,...,M$) defined in a set of $M$ probing locations belonging
to a \emph{RoI} $\Omega\subset\mathcal{D}$, find the optimal set
of array excitation phases $\underline{\beta}^{opt}=\left\{ \beta_{n}^{opt};\,\  n=1,...,N\right\} $
such that \begin{equation}
\underline{\beta}^{opt}=\arg\left\{ \min_{\underline{\beta}}\Phi\left(\underline{\beta}\right)\right\} \label{eq:Solution}\end{equation}
where\begin{equation}
\Phi\left(\underline{\beta}\right)=\frac{1}{M}\sum_{m=1}^{M}\frac{\left|P_{rx}\left(\left.\underline{r}_{m}\right|\underline{\beta}\right)-P_{rx}^{tar}\left(\underline{r}_{m}\right)\right|}{\left|P_{rx}^{tar}\left(\underline{r}_{m}\right)\right|}\label{eq:Cost-function-PSO}\end{equation}
is the cost function quantifying the mismatch between the received
power and the target one within $\Omega$.
\end{quotation}

\section{\emph{OSSP} Solution Method \label{sub:OSSP-Solution-Methodology}}

To effectively explore the $N$-dimensional solution space by sampling
the non-convex landscape of (\ref{eq:Cost-function-PSO}) to find
the global optimal solution of the \emph{OSSP}, $\underline{\beta}^{opt}$,
nature-inspired evolutionary algorithms (\emph{EA}s) are promising
candidates owing to their hill-climbing features and the avoidance
of the differentiation of the mismatch cost function (\ref{eq:Cost-function-PSO})
for the search. More specifically, due to the real-valued nature of
the \emph{DoF}s at hand (\ref{eq:DoFs}) as well as the highly non-linear/multi-modal
nature of (\ref{eq:Cost-function-PSO}), the Particle Swarm Optimization
(\emph{PSO}) method \cite{Robinson 2004}-\cite{Goudos 2018} is here
exploited as the core of the optimization engine for solving (\ref{eq:Solution}).
However, a {}``bare'' integration of the \emph{PSO} with a forward
\emph{EM} simulator would imply an overall computational time equal
to\begin{equation}
\Delta t_{PSO}=\left(K\times I\right)\times\Delta t_{sim}\label{eq:}\end{equation}
where $K\propto N$ is the number of particles (trial solutions) evaluated
at each $i$-th ($i=1,...,I$) iteration, $I$ being the total number
of iterations, while $\Delta t_{sim}$ is the \emph{CPU} time for
a single evaluation of (\ref{eq:Cost-function-PSO}). As for this
latter, it is worth noticing that the prediction of the received power
distribution within $\Omega$ for any trial set of the excitation
phases ($\underline{\beta}$) requires the simulation of the \emph{BTS}
array radiating within the complex scattering scenario $\mathcal{D}$.
This leads to an impractical computational burden despite the exploitation
of efficient forward solvers based on ray-tracing (\emph{RT}) techniques
instead of using full-wave methods \cite{Benoni 2022}, which are
fully hindered by the large-scale of the urban scenario at hand. 

\noindent In order to reduce the computational load of the optimization
process to a manageable amount, the concept of \emph{Embedded-plus-Environment
Pattern} (\emph{EPEP}) is introduced. Indeed, such an approach allows
one to perform an fast and faithful prediction of the received power
within $\Omega$ without recurring to iterated/time-consuming \emph{RT}-based
simulations during the optimization. Under the assumption that the
\emph{EM} waves radiated by the \emph{BTS} propagate in linear materials,
which is a safe assumption for common/naturally-occurring media \cite{Kelley 1993},
it is possible to express the total electric field radiated by the
array with phase vector $\underline{\beta}$, $E_{\gamma}\left(\left.\underline{r}\right|\underline{\beta}\right)$
$\underline{r}\in\mathcal{D}$, as the linear superposition of the
\emph{EM} field distributions obtained by making each $n$-th ($n=1,...,N$)
embedded element within the array radiate individually within the
complex scattering scenario. Mathematically, it turns out that \begin{equation}
E_{\gamma}\left(\left.\underline{r}\right|\underline{\beta}\right)=\sum_{n=1}^{N}\xi\times\exp\left(j\beta_{n}\right)\times E_{\gamma}^{\left(n\right)}\left(\left.\underline{r}\right|\mathcal{D}\right)\label{eq:total-field-from-EPEP}\end{equation}
where $E_{\gamma}^{\left(n\right)}\left(\left.\underline{r}\right|\mathcal{D}\right)$
is the $\gamma$-th ($\gamma=\left\{ x,y,z\right\} $) component of
the \emph{EPEP} of the $n$-th ($n=1,...,N$) radiator ($\underline{E}^{\left(n\right)}\left(\left.\underline{r}\right|\mathcal{D}\right)=\sum_{\gamma=\left\{ x,y,z\right\} }E_{\gamma}^{\left(n\right)}\left(\left.\underline{r}\right|\mathcal{D}\right)\underline{u}_{\gamma}$),
which is computed by setting the array excitations as follows\begin{equation}
w_{q}=\left\{ \begin{array}{ll}
1.0 & \mathrm{if}\, q=n\\
0.0 & \mathrm{otherwise}\end{array}\right.\label{eq:EPEP-Excitations}\end{equation}
and considering the $\left(N-1\right)$ elements with null-excitations
connected to matched loads. Since each $n$-th ($n=1,...,N$) \emph{EPEP},
$\underline{E}^{\left(n\right)}\left(\left.\underline{r}\right|\mathcal{D}\right)$,
is computed by simulating the whole array in the complex scattering
environment $\mathcal{D}$, it turns out that $E_{\gamma}^{\left(n\right)}\left(\left.\underline{r}\right|\mathcal{D}\right)$
($\gamma=\left\{ x,y,z\right\} $) accounts not only for the \emph{EM}
interactions between neighboring elements (e.g., their mutual coupling
\cite{Kelley 1993}-\cite{Aumann 1989}), but also those introduced
by the surrounding environment.

\noindent Thanks to the \emph{EPEP} method, the iterative numerical
prediction of the power coverage in $\Omega$ during the optimization
loop is avoided. As a matter of fact, for any given scenario $\mathcal{D}$
and \emph{BTS} array setup, it is enough to run $N$ \emph{RT}-based
simulations for \emph{off-line} building a database of \emph{EPEP}s
before entering the minimization process, where the received power
associated to any trial guess solution $\underline{\beta}$ can be
analytically computed by inputting the result of (\ref{eq:total-field-from-EPEP})
into (\ref{eq:received-power}) without further using the (time-consuming)
forward simulator. As a result, the time saving with respect to the
{}``bare'' approach in the overall synthesis process amounts to\begin{equation}
\Delta t_{sav}=\frac{\left(K\times I\right)-N}{\left(K\times I\right)}.\label{eq:}\end{equation}

\noindent The proposed \emph{OSS} method then consists of the following
procedural steps:

\begin{enumerate}
\item \emph{Input Phase} - Define the \emph{BTS} array aperture, $\Psi$,
and the number of array elements, $N$. Compute the magnitude of the
excitations, $\xi$, according to (\ref{eq:excitation-magnitude}).
Define the complex scattering scenario $\mathcal{D}$ and discretize
the selected \emph{RoI} $\Omega$ into $M$ uniformly-spaced probing
locations. Input the target received power distribution in $\Omega$
(i.e., $P_{rx}^{tar}\left(\underline{r}_{m}\right)$; $m=1,...,M$);
\item \emph{EPEP}s \emph{Database Computation} - Run $N$ \emph{RT}-based
simulations of the array lying within the scenario $\mathcal{D}$
by setting each time ($n=1,...,N$) the excitations according to (\ref{eq:EPEP-Excitations})
to fill the database $\mathbb{E}$ of the \emph{EPEP}s\begin{equation}
\mathbb{E}=\left\{ \left[\underline{E}^{\left(n\right)}\left(\left.\underline{r}_{m}\right|\mathcal{D}\right);\, m=1,...,M\right];\, n=1,...,N\right\} ;\label{eq:}\end{equation}

\item \emph{Synthesis Initialization} ($i=0$) - Randomly initialize a swarm
of $K$ particles/trial-solution $\mathbb{B}_{0}=\left\{ \underline{\beta}_{0}^{\left(k\right)};\, k=1,...,K\right\} $
and the associated velocities $\mathbb{V}_{0}=\left\{ \underline{v}_{0}^{\left(k\right)};\, k=1,...,K\right\} $.
Initialize the personal best position vector $\mathbb{T}_{0}$ by
setting its $k$-th ($k=1,...,K$) entry to $\underline{t}_{0}^{\left(k\right)}=\underline{\beta}_{0}^{\left(k\right)}$;
\item \emph{Synthesis Loop} ($i=0,...,I$) 

\begin{enumerate}
\item Use (\ref{eq:total-field-from-EPEP}) and (\ref{eq:received-power})
to analytically compute the received power distribution in $\Omega$
for each $k$-th ($k=1,...,K$) trial solution of the current-iteration
swarm $\mathbb{B}_{i}=\left\{ \underline{\beta}_{i}^{\left(k\right)};\, k=1,...,K\right\} $;
\item Compute the corresponding cost function values, $\Phi\left\{ \underline{\beta}_{i}^{\left(k\right)}\right\} $($k=1,...,K$),
according to (\ref{eq:Cost-function-PSO});
\item Update the global best at the current $i$-th iteration as $\underline{\beta}_{i}^{opt}=\arg\left[\min_{k=1,...,K;\, j=0,...,i}\Phi\left\{ \underline{\beta}_{j}^{\left(k\right)}\right\} \right]$.
If $i>0$, then update each $k$-th ($k=1,...,K$) personal best,
$\underline{t}_{i}^{\left(k\right)}\leftarrow\underline{\beta}_{i}^{\left(k\right)}$,
when $\Phi\left\{ \underline{\beta}_{i}^{\left(k\right)}\right\} <\Phi\left\{ \underline{t}_{i-1}^{\left(k\right)}\right\} $,
otherwise let $\underline{t}_{i}^{\left(k\right)}\leftarrow\underline{t}_{i-1}^{\left(k\right)}$;
\item Stop the optimization if $i=I$ or if the following stagnation condition
holds true \cite{Rocca 2009}\begin{equation}
\frac{\left|W\times\Phi\left\{ \underline{\beta}_{i}^{opt}\right\} -\sum_{j=1}^{W}\Phi\left\{ \underline{\beta}_{i-j}^{opt}\right\} \right|}{\Phi\left\{ \underline{\beta}_{i}^{opt}\right\} }\le\rho,\label{eq:Stagnation-check}\end{equation}
 $\rho$ and $W$ being a fixed threshold and a user-defined number
of iterations, respectively, and output $\underline{\beta}^{opt}=\underline{\beta}_{i}^{opt}$; 
\item Update the velocities, $\mathbb{V}_{i+1}\leftarrow\mathbb{V}_{i}$,
and the positions of each particle, $\mathbb{B}_{i+1}\leftarrow\mathbb{B}_{i}$,
according to the \emph{PSO} update rules with a fixed inertial weight
$\varpi$ and cognitive/social acceleration coefficients $\mathcal{C}_{1}/\mathcal{C}_{2}>0$
\cite{Rocca 2009}. Then, let $i\leftarrow\left(i+1\right)$ and repeat
from step 4(\emph{a}).
\end{enumerate}
\end{enumerate}

\section{Numerical Results\label{sec:Numerical-Results}}

The goal of this Section is two-fold. First, the \emph{EPEP} approach
for the computationally-fast/reliable evaluation of the received power
distribution within large-scale complex-scattering scenarios is validated
(Sect. \ref{sub:Embedded-plus-environment-Approach-Validation}).
Afterward, practical examples of the \emph{OSS}-enabled implementation
of \emph{SEME}s in realistic outdoor areas are reported to assess
the capabilities and the potentialities of the proposed method (Sect.
\ref{sub:Opportunistic-Sources-Synthesis}).

\noindent In the following, the primary source $\Psi$ is assumed
to be a planar array of $N=\left(4\times8\right)=32$ elements on
the $\left(x,\, z\right)$ plane, placed at height $z_{\Psi}=20$
{[}m{]} above the ground ($z=0$ {[}m{]}) with a mechanical down-tilt
of $\theta_{\Psi}=2$ {[}deg{]} \cite{Benoni 2022} {[}Fig. 1(\emph{b}){]}.
Such a planar aperture is filled by $d_{x}=d_{z}=\lambda/2$-spaced
radiators that are numerically modelled in the \emph{ANSYS HFSS} simulator
\cite{ANSYS HFSS} as dual-polarized (slant-45) slot-fed square patches
working at a central frequency of $f=3.5$ {[}GHz{]}%
\footnote{\noindent For symmetry reasons, only the $+45$ {[}deg{]}-slanted
polarization is analyzed in the following, the performance for the
$-45$ {[}deg{]} polarization being equivalent.%
} \cite{Benoni 2022} (Fig. 2). Unless otherwise stated, the maximum
radiated power by the array has been set to $\zeta^{\max}=20$ {[}W{]}
($\rightarrow\xi=8.63$) \cite{3G Americas}-\cite{Huo 2017}. As
for the definition of the \emph{RoI}s, planar surfaces parallel to
the $\left(x,\, y\right)$ plane have been uniformly sampled by spacing
the $M$ probing locations along $x$ and $y$ with step $\Delta_{x}=\Delta_{y}=5$
{[}m{]} at a fixed height $z_{\Omega}=h=1.5$ {[}m{]}%
\footnote{\noindent The height of the \emph{RoI}s has been set to the average
height of mobile user equipment in realistic scenarios \cite{Benoni 2022}.%
} {[}Fig. 1(\emph{a}){]}. Moreover, the received power distribution
in $\mathcal{D}$ has been predicted with the \emph{3D-RT} solver
of the \emph{Altair WINPROP} software suite \cite{Altair Winprop}
to model all the interactions between the primary source $\Psi$ and
the real-world urban scenario, extracted from the OpenStreetMap (\emph{OSM})
database \cite{OpenStreetMap}, where the buildings have been simulated
with external walls of thickness $t_{w}=0.3$ {[}m{]} made of concrete
with $\varepsilon_{r}=6$ and $\sigma=0.136$ {[}S/m{]} according
to \cite{Altair Winprop}\cite{Wong 2021}\cite{Zadhoush 2021}.

\noindent Finally, the target power distribution in (\ref{eq:Cost-function-PSO})
(i.e., $P_{rx}^{tar}\left(\underline{r}_{m}\right)$; $m=1,...,M$),
has been set to that generated within $\Omega$ by an \emph{ideal}
source $\Psi'$ modeled as a larger array of $N'>N$ elements with
uniform excitations (i.e., $\underline{w}'=\left\{ w_{n}^{'}=\xi;\, n=1,...,N'\right\} $),
placed in the same position of $\Psi$ (i.e., $\underline{r}_{\Psi'}=\underline{r}_{\Psi}$),
but radiating within an ideal scenario $\mathcal{D}'$ without any
obstacle (i.e., {}``free-space'' propagation conditions). Moreover,
the ideal array $\Psi'$ has been mechanically tilted towards the
\emph{RoI} barycenter, $\underline{r}_{\Omega}$, by setting the azimuth
direction $\varphi_{\Psi}^{'}$ as in Fig. 1(\emph{c}) to maximize
the power transfer, being $\theta_{\Psi'}=\theta_{\Psi}$.

\subsection{\emph{EPEP} Approach Validation\label{sub:Embedded-plus-environment-Approach-Validation}}

To provide a numerical validation of the \emph{EPEP} approach for
the coverage prediction (Sect. \ref{sub:OSSP-Solution-Methodology}),
a $300\times350$ {[}$m^{2}${]} portion of the municipality of Orgiano
(Vicenza, Italy) has been chosen as the propagation scenario $\mathcal{D}$
(Fig. 3). The \emph{BTS} array has been located at the position $\left(x_{\Psi},\, y_{\Psi}\right)=\left(231,\,88\right)$
{[}m{]} with azimuth orientation $\varphi_{\Psi}=0$ {[}deg{]} (Fig.
3). Figure 4 shows a representative subset of the $N=32$ \emph{EPEP}s
obtained by simulating each $n$-th ($n=1,...,N$) embedded element
of the array that radiates in the domain $\mathcal{D}$. In particular,
the $x-$ (left column - Fig. 4), $y-$ (middle column - Fig. 4),
and $z-$ (right column - Fig. 4) components of the electric field
amplitude related to the $n=1$ {[}Figs. 4(\emph{a})-4(\emph{c}){]},
the $n=12$ {[}Figs. 4(\emph{d})-4(\emph{f}){]}, and the $n=32$ {[}Figs.
4(\emph{g})-4(\emph{i}){]} (i.e., two corner elements and one central
element - Fig. 2) array elements are reported. To prove that such
field distributions form a basis for the received power distribution
in $\mathcal{D}$ for any set of the array excitations, Figure 5 compares
the received power distribution, $P_{rx}\left(x,y\right)$, $\left(x,y\right)\in\mathcal{D}$,
computed as the linear combination of the \emph{EPEP}s according to
(\ref{eq:total-field-from-EPEP})(\ref{eq:received-power}) with that
obtained by directly simulating the array with all elements contemporarily
fed, $\widehat{P}_{rx}\left(x,y\right)$, $\left(x,y\right)\in\mathcal{D}$.
More specifically, three benchmark scenarios have been considered
for the $N$ phase coefficients of the \emph{BTS} array: (\emph{i})
\emph{uniform}\begin{equation}
\underline{\beta}^{uni}=\left\{ \beta_{n}^{uni}=0;\,\,\, n=1,...,N\right\} \label{eq:}\end{equation}
 {[}Fig. 5(\emph{a}) vs. Fig. 5(\emph{d}){]}, (\emph{ii}) $\left(\theta_{s},\,\varphi_{s}\right)=\left(95,-60\right)$
{[}deg{]} \emph{beam-steering} \cite{Li 2005}\begin{equation}
\underline{\beta}^{ste}\left(\theta_{s},\,\varphi_{s}\right)=\left\{ \beta_{n}^{ste}\left(\theta_{s},\,\varphi_{s}\right)=-2\pi\times\left[x_{n}\times\sin\left(\theta_{s}\right)\times\cos\left(\varphi_{s}\right)+z_{n}\times\cos\left(\theta_{s}\right)\right];\,\,\, n=1,...,N\right\} ,\label{eq:standard-steering-phases}\end{equation}
 $\left(x_{n},\, z_{n}\right)$ being the barycenter of the $n$-th
($n=1,...,N$) array element {[}Fig. 5(\emph{b}) vs. Fig. 5(\emph{e}){]},
and (\emph{iii}) \emph{random} \begin{equation}
\underline{\beta}^{rnd}=\left\{ \beta_{n}^{rnd}=\mathcal{R}\left\{ 0,\,2\pi\right\} ;\,\,\, n=1,...,N\right\} ,\label{eq:}\end{equation}
 $\mathcal{R}\left\{ a,\, b\right\} $ being a uniformly-distributed
random number in the interval $\left[a,\, b\right]$ {[}Fig. 5(\emph{c})
vs. Fig. 5(\emph{f}){]}. Pictorially, the analytically-computed received
power distribution starting from the \emph{EPEP}s quite faithfully
matches that from the complete array simulation. This is further pointed
out by the difference maps ($\Delta\widehat{P}_{rx}\left(\underline{r}\right)\triangleq\left|\widehat{P}_{rx}\left(\underline{r}\right)-P_{rx}\left(\underline{r}\right)\right|$)
in Figs. 5(\emph{g})-5(\emph{i}). As it can be observed, there are
only some negligible deviations between $\widehat{P}_{rx}\left(\underline{r}\right)$
and $P_{rx}\left(\underline{r}\right)$ mainly in correspondence with
the regions far from the \emph{BTS} where the received power is generally
so low (i.e., $\widehat{P}_{rx}\left(\underline{r}\right)<-70$ {[}dBm{]})
that the optical approximation errors introduced by the \emph{RT}
solver, start to appear. Quantitatively, the average difference between
$P_{rx}\left(\underline{r}\right)$ and $\widehat{P}_{rx}\left(\underline{r}\right)$
is very small and equal to $\left.\mathrm{avg}_{\underline{r}\in\mathcal{D}}\left\{ \Delta\widehat{P}_{rx}\right\} \right|_{uni}=0.20$
{[}dBm{]} {[}Fig. 5(\emph{g}){]}, $\left.\mathrm{avg}_{\underline{r}\in\mathcal{D}}\left\{ \Delta\widehat{P}_{rx}\right\} \right|_{ste}=0.15$
{[}dBm{]} {[}Fig. 5(\emph{h}){]}, and $\left.\mathrm{avg}_{\underline{r}\in\mathcal{D}}\left\{ \Delta\widehat{P}_{rx}\right\} \right|_{rnd}=0.10$
{[}dBm{]} {[}Fig. 5(\emph{i}){]} for the uniform, steered, and random
excitations, respectively.

\subsection{\emph{OSS} Method Validation\label{sub:Opportunistic-Sources-Synthesis}}

The first assessment of the proposed \emph{OSS} method deals with
a square \emph{RoI} $\Omega$ in the {}``Orgiano'' scenario of extension
$\mathcal{A}\left(\Omega\right)=\left(35\times35\right)=1225$ {[}$\mathrm{m}^{2}${]}
{[}$\rightarrow M=\left(7\times7\right)=49${]} centered in $\underline{r}_{\Omega}=\left(122.5,\,262.5\right)$
{[}m{]}. An ideal array $\Psi'$ with $N'=\left(4\times9\right)=36$
elements (i.e., $\Delta N=\frac{N'-N}{N}=+12\%$ more elements than
$\Psi$) has been chosen to define the target received power {[}Fig.
6(\emph{a}){]}. The control parameters of the \emph{OSS} method have
been set to \cite{Rocca 2009}: $K=\left(2\times N\right)=64$, $I=10^{3}$,
$W=10^{2}$, $\rho=10^{-3}$, $\varpi=0.4$, and $\mathcal{C}_{1}/\mathcal{C}_{2}=2.0$.
As a result, the time saving over a {}``bare'' optimization with
the same parameters, but without relying on the off-line generation
of the \emph{EPEP}s database $\mathbb{E}$, %
\footnote{\noindent The \emph{CPU} cost required to build the database $\mathbb{E}$
of $N=32$ \emph{EPEP}s and to perform the synthesis are equal to
$\Delta t_{\mathbb{E}}\approx1.44\times10^{3}$ {[}sec{]} and $\Delta t_{OSS}\approx6.0$
{[}sec{]}, respectively, on a standard computer equipped with Intel
Core i5 \emph{CPU} @ 1.60 {[}GHz{]} with 16 {[}GB{]} of \emph{RAM}
memory.%
} turned out to be of $\Delta t_{sav}=99.95$ {[}\%{]}.

\noindent Figure 7 reports an outcome of the optimization process
in terms of the evolution of the normalized cost function\begin{equation}
\hat{\Phi}\left(\underline{\beta}\right)\triangleq\frac{\Phi\left(\underline{\beta}\right)}{\Phi\left(\underline{\beta}^{uni}\right)},\label{eq:Normalized-cost}\end{equation}
 $\Phi\left(\underline{\beta}^{uni}\right)$ being the cost function
computed for the reference scenario with uniform excitations {[}Fig.
6(\emph{b}){]}. As it can be observed, the \emph{OSS} method has found
the convergence solution $\underline{\beta}^{opt}$ after $I_{conv}=400$
iterations when the stagnation condition (\ref{eq:Stagnation-check})
holds true, the improvement with respect to the case of uniform excitations
(i.e., $\hat{\Phi}\left(\underline{\beta}^{opt}\right)=69.5\%$) being
$\Delta\hat{\Phi}=30.5\%$ ($\Delta\hat{\Phi}\triangleq\left[\hat{\Phi}\left(\underline{\beta}^{uni}\right)-\hat{\Phi}\left(\underline{\beta}^{opt}\right)\right]$).

\noindent To give more insights on the synthesized solution, the entries
of the optimized phase vector $\underline{\beta}^{opt}$ are reported
in Fig. 8(\emph{a}), while the corresponding normalized power pattern
is shown in Fig. 8(\emph{b}). As it can be observed, the synthesized
excitations {[}Fig. 8(\emph{a}){]} shape the \emph{BTS} pattern in
an unconventional way since they generate two main radiation beams
towards the angular directions $\left(\theta_{1},\varphi_{1}\right)\approx\left(90.5,\,97.5\right)$
{[}deg{]} {[}$\rightarrow\left(u=-1.3\times10^{-1},\, v=-8.7\times10^{-3}\right)${]}
and $\left(\theta_{2},\varphi_{2}\right)\approx\left(90.5,\,139.5\right)$
{[}deg{]} {[}$\rightarrow\left(u=-7.6\times10^{-1},\, v=-8.7\times10^{-3}\right)${]},
respectively {[}Fig. 8(\emph{b}){]}. By analyzing the corresponding
received power distribution $P_{rx}^{opt}\left(\underline{r}\right)=P_{rx}\left(\left.\underline{r}\right|\underline{\beta}^{opt}\right)$
in Fig. 9(\emph{a}), where the paths of the highest-energy rays propagating
from $\underline{r}_{\Psi}$ to $\underline{r}_{\Omega}$ are also
shown, it turns out that such directions correspond to the two main
propagation trajectories allowing to circumvent the large rectangular
building obstructing the \emph{LOS} between the \emph{BTS} and the
\emph{RoI}. This is a clear indication that the synthesized primary
source $\Psi$ opportunistically exploits the surrounding buildings
to enhance the connectivity within a specific region otherwise suffering
from the negative effects of \emph{NLOS}/shadowing. 

\noindent Even more interesting, the \emph{OSS} solution remarkably
outperforms that performing a standard beam-steering of the \emph{BTS}
pattern towards the \emph{RoI} barycenter $\underline{r}_{\Omega}$.
Indeed, the set of the array excitations according to (\ref{eq:standard-steering-phases})
with $\left(\theta_{s},\,\varphi_{s}\right)=\left(95.1,\,121.9\right)$
{[}deg{]} in Fig. 8(\emph{a}) (green dashed curve), which affords
the \emph{FF} pattern in Fig. 8(\emph{d}), results in a \emph{RoI}
coverage even worse than that of the uniform case ($\hat{\Phi}\left(\underline{\beta}^{ste}\right)=119.8\%$)
because of the occlusion of the direct path between the source and
the receivers {[}Fig. 6(\emph{c}){]}. To better illustrate such outcomes,
the plots of the absolute target power mismatch in the \emph{RoI},
$\left|\Delta P_{rx}^{tar}\left(\underline{r}\right)\right|$ $\underline{r}\in\Omega$
($\left|\Delta P_{rx}^{tar}\left(\underline{r}\right)\right|\triangleq\left|P_{rx}\left(\underline{r}\right)-P_{rx}^{tar}\left(\underline{r}\right)\right|$)
are shown in Figs. 9(\emph{c})-9(\emph{e}). As it can be inferred,
the synthesized excitation vector $\underline{\beta}^{opt}$ provides
a better matching with the received power level of the ideal source
$\Psi'$ than the other two {}``conventional'' solutions {[}Fig.
9(\emph{c}) vs. Figs. 9(\emph{d})-9(\emph{e}){]}. Quantitatively,
it turns out that the average power improvement within the \emph{RoI}
is of about $\left(\mathrm{avg}_{\underline{r}\in\Omega}\left\{ P_{rx}^{opt}\left(\underline{r}\right)\right\} -\mathrm{avg}_{\underline{r}\in\Omega}\left\{ P_{rx}^{uni}\left(\underline{r}\right)\right\} \right)=2.2$
{[}dBm{]} {[}Fig. 9(\emph{b}){]} and $\left(\mathrm{avg}_{\underline{r}\in\Omega}\left\{ P_{rx}^{opt}\left(\underline{r}\right)\right\} -\mathrm{avg}_{\underline{r}\in\Omega}\left\{ P_{rx}^{ste}\left(\underline{r}\right)\right\} \right)=15.2$
{[}dBm{]} {[}Fig. 9(\emph{b}){]}, respectively (Tab. I). 

\noindent From a practical point of view, it is worth remarking that
the \emph{OSS} solution considerably increases the minimum value of
the received power within $\Omega$ and, in turn, the overall \emph{QoS}
to the end-users since $\left(\min_{\underline{r}\in\Omega}\left\{ P_{rx}^{opt}\left(\underline{r}\right)\right\} -\right.$
$\left.\min_{\underline{r}\in\Omega}\left\{ P_{rx}^{uni}\left(\underline{r}\right)\right\} \right)=14.7$
{[}dBm{]} and $\left(\min_{\underline{r}\in\Omega}\left\{ P_{rx}^{opt}\left(\underline{r}\right)\right\} -\right.$
$\left.\min_{\underline{r}\in\Omega}\left\{ P_{rx}^{ste}\left(\underline{r}\right)\right\} \right)=14.1$
{[}dBm{]}, respectively (Tab. I). For completeness, the plot of the
difference map $\Delta P_{rx}^{opt}\left(\underline{r}\right)$($\Delta P_{rx}^{opt}\left(\underline{r}\right)\triangleq\left[P_{rx}^{opt}\left(\underline{r}\right)-P_{rx}\left(\underline{r}\right)\right]$)
is shown in Fig. 10(\emph{a}) {[}$P_{rx}\left(\underline{r}\right)=P_{rx}^{uni}\left(\underline{r}\right)${]}
and Fig. 10(\emph{b}) {[}$P_{rx}\left(\underline{r}\right)=P_{rx}^{ste}\left(\underline{r}\right)${]}
to point out that main impact of the \emph{OSS} method is in the bottom-left
portion of the \emph{RoI} {[}see Fig. 9(\emph{c}) vs. Figs. 9(\emph{d})-9(\emph{e}){]},
which is an almost-blind area in both the uniform and the steered
cases.

\noindent The second test case is concerned with a larger outdoor
area ($1000\times700$ {[}$m^{2}${]}) located within the hinterland
region of the city of Padova in Italy (Fig. 11). Figure 12(\emph{a})
shows the received power distribution generated by $\Psi'$ in free-space
conditions, while the \emph{RoI} $\Omega_{1}$ has an area of $\mathcal{A}\left(\Omega_{1}\right)=\left(80\times80\right)=6400$
{[}$\mathrm{m}^{2}${]} {[}$\rightarrow M=\left(16\times16\right)=256${]}
and it is centered at $\underline{r}_{\Omega_{1}}=\left(785,\,340\right)$
{[}m{]}. In such a scenario, the optimal solution found by the \emph{OSS}
process improves of about $\Delta\hat{\Phi}=48.9\%$ that with the
uniform excitations {[}Fig. 12(\emph{b}){]} by almost halving the
mismatch with respect to the target distribution. Analogously, it
significantly differs from that steering the beam towards the center
of the \emph{RoI} {[}Fig. 13(\emph{a}){]} by affording a pattern with
two main lobes occurring at $\left(\theta_{1},\varphi_{1}\right)\approx\left(90,\ 78.5\right)$
{[}deg{]} {[}$\rightarrow\left(u=2.0\times10^{-1},\, v=0.0\right)${]}
and $\left(\theta_{2},\varphi_{2}\right)\approx\left(90,\ 131\right)$
{[}deg{]} {[}$\rightarrow\left(u=-6.6\times10^{-1},\, v=0.0\right)${]},
respectively {[}Fig 13(\emph{b}){]}. Once again, the corresponding
received power distribution shows that such radiation peaks correspond
to {}``convenient'' directions allowing the \emph{BTS} to exploit
at best the surrounding obstacles to scatter the field towards the
\emph{RoI} {[}Fig. 14(\emph{a}){]}. For comparison purposes, the excitations
to directly steer the beam pattern towards the \emph{RoI} {[}Fig.
12(\emph{c}){]} provide a much worse coverage (i.e., $\hat{\Phi}\left(\underline{\beta}^{ste}\right)=112.5\%$).
Once again, these outcomes are confirmed by the maps of $\left|\Delta P_{rx}^{tar}\left(\underline{r}\right)\right|$
{[}Figs. 14(\emph{c})-14(\emph{e}){]} and $\Delta P_{rx}^{opt}\left(\underline{r}\right)$
(Fig. 15). Quantitatively, the average power improvement by the \emph{OSS}
is equal to $\left(\mathrm{avg}_{\underline{r}\in\Omega}\left\{ P_{rx}^{opt}\left(\underline{r}\right)\right\} -\mathrm{avg}_{\underline{r}\in\Omega}\left\{ P_{rx}^{uni}\left(\underline{r}\right)\right\} \right)=7.9$
{[}dBm{]} (Tab. I) and $\left(\mathrm{avg}_{\underline{r}\in\Omega}\left\{ P_{rx}^{opt}\left(\underline{r}\right)\right\} -\mathrm{avg}_{\underline{r}\in\Omega}\left\{ P_{rx}^{ste}\left(\underline{r}\right)\right\} \right)=11.9$
{[}dBm{]} (Tab. I). Furthermore, the minimum power level within the
\emph{RoI} $\Omega_{1}$ is raised by $\left(\min_{\underline{r}\in\Omega}\left\{ P_{rx}^{opt}\left(\underline{r}\right)\right\} -\min_{\underline{r}\in\Omega}\left\{ P_{rx}^{uni}\left(\underline{r}\right)\right\} \right)=17.8$
{[}dBm{]} and $\left(\min_{\underline{r}\in\Omega}\left\{ P_{rx}^{opt}\left(\underline{r}\right)\right\} -\min_{\underline{r}\in\Omega}\left\{ P_{rx}^{ste}\left(\underline{r}\right)\right\} \right)=12.6$
{[}dBm{]}, respectively (Tab. I).

\noindent Let us now investigate on the effect on the coverage performance
of the increasing of the maximum radiated power $\zeta^{\max}$ of
the \emph{BTS} array. Towards this end, a set of optimizations has
been run by scaling the magnitude of the excitations according to
the following rule\begin{equation}
\alpha_{n}=\delta\times\xi;\,\,\, n=1,...,N\label{eq:}\end{equation}
and varying the scaling factor $\delta$ within the range $\delta\in\left[1,\,5\right]$
so that the resulting maximum radiated power turns out to lie in the
interval $\zeta^{\max}\in\left[20,\,500\right]$ {[}W{]} ($\zeta^{\max}=\delta^{2}\times20$
{[}W{]}). 

\noindent The plot of the cost function vs. $\zeta^{\max}$ in Fig.
16 shows that there is a progressive improvement of the matching when
increasing $\zeta^{\max}$. As a matter of fact, the availability
of more transmitting power allows the \emph{BTS} to {}``smartly''
radiate it along different directions and at a greater distance to
involve more obstacles of the surrounding environment for better matching
of the target power in the \emph{RoI}. Such a consideration is supported
by the plot of the synthesized patterns (Fig. 17) that become more
unconventional (e.g., multi-beams) as the available power increases.
For completeness, the coverage maps are given in Fig. 18, while the
matching with the target distribution is highlighted by the power
maps in Fig. 19, which statistics values are given in Tab. II. 

\noindent Finally, the last test cases are aimed at assessing the
flexibility of the proposed method and its effectiveness when dealing
with different locations of the \emph{RoI}. Towards this purpose,
the \emph{OSS} optimizations have been performed by considering other
two different \emph{RoI}s (i.e., $\Omega_{2}$ or $\Omega_{3}$) within
the same scenario of the previous example, but centered at either
$\underline{r}_{\Omega_{2}}=\left(875,\,320\right)$ {[}m{]} or $\underline{r}_{\Omega_{3}}=\left(785,\,365\right)$
{[}m{]} and having a support $\mathcal{A}\left(\Omega_{2}\right)=\left(80\times80\right)=6400$
{[}$\mathrm{m}^{2}${]} ($\rightarrow M=256$) or $\mathcal{A}\left(\Omega_{3}\right)=\left(50\times50\right)=2500$
{[}$\mathrm{m}^{2}${]} ($\rightarrow M=100$), respectively.

\noindent The \emph{OSS} results are illustrated in Fig. 20. As it
can be inferred, the \emph{BTS} modifies its excitations to radiate
towards different directions depending on the \emph{RoI} at hand {[}Fig.
20(\emph{a}) vs. Fig. 20(\emph{b}){]} for exploiting at best the scattering
phenomena generated with the surrounding environment {[}Fig. 20(\emph{c})
vs. Fig. 20(\emph{d}){]}. As a matter of fact, unlike the solution
found for $\Omega_{1}$ {[}Fig. 14(\emph{a}){]}, where there is a
split of the radiated power along two main directions, the best coverage
of $\Omega_{2}$ or $\Omega_{3}$ corresponds to different steerings
of a single main beam towards $\left(\theta,\,\varphi\right)\approx\left(90,\ 83\right)$
{[}deg{]} {[}$\rightarrow\left(u=1.2\times10^{-1},\, v=0.0\right)${]}
{[}Fig. 20(\emph{a}){]} or $\left(\theta,\varphi\right)\approx\left(90,\ 128\right)$
{[}deg{]} {[}$\rightarrow\left(u=-6.2\times10^{-1},\, v=0.0\right)${]}
{[}Fig. 20(\emph{b}){]}. In both cases, there is still a good matching
with the target power distribution {[}Figs. 20(\emph{e})-20(\emph{f}){]}
as well as a remarkable increase of the average and minimum received
power with respect to conventional (uniform/steered) solutions (Tab.
I).

\section{Conclusions\label{sec:Conclusions}}

Within the emerging \emph{SEME} framework, an innovative strategy
for affording a target power distribution over a user-defined \emph{RoI}
has been presented. Rather than relying on the optimal design/planning
of suitable field manipulating devices (e.g., \emph{EMS}s) to arbitrarily
tailor the propagation features of the environment, an optimization
strategy has been proposed to design a {}``smart \emph{BTS}'' able
to reconfigure itself for opportunistically exploiting the surrounding
environment. Such an approach is based on the concept of \emph{EPEP}
that allows one to perform a fast and faithful prediction of the received
power within $\Omega$ without recurring to iterated/time-consuming
\emph{RT}-based simulations during the optimization.

\noindent Selected numerical results, concerned with real-world propagation
scenarios, have been reported and discussed to assess the capabilities
and the potentialities of the proposed \emph{OSS} method. 

\noindent Future works, beyond the scope of this manuscript, will
be aimed at extending the proposed approach to deal with (\emph{a})
different scenarios/sources (e.g., indoor scenarios where the primary
source is a Wi-Fi access points) and (\emph{b}) alternative definitions
of the cost function to take into account specific system requirements
and user needs. Moreover, the integration of the proposed strategy
with other \emph{SEME} technologies and methodologies will be the
subject of future research tracks.

\section*{Acknowledgements}

This work benefited from the networking activities carried out within
the Project {}``SPEED'' (Grant No. 6721001) funded by National Science
Foundation of China under the Chang-Jiang Visiting Professorship Program,
the Project {}``AURORA - Smart Materials for Ubiquitous Energy Harvesting,
Storage, and Delivery in Next Generation Sustainable Environments''
funded by the Italian Ministry for Universities and Research within
the PRIN-PNRR 2022 Program (CUP: E53D23014760001), the Project {}``Smart
ElectroMagnetic Environment in TrentiNo - SEME@TN'' funded by the
Autonomous Province of Trento (CUP: C63C22000720003), the Project
DICAM-EXC (Grant L232/2016) funded by the Italian Ministry of Education,
Universities and Research (MUR) within the {}``Departments of Excellence
2023-2027'' Program (CUP: E63C22003880001), and the Project {}``ICSC
National Centre for HPC, Big Data and Quantum Computing (CN HPC)''
funded by the European Union - NextGenerationEU within the PNRR Program
(CUP: E63C22000970007). Views and opinions expressed are however those
of the author(s) only and do not necessarily reflect those of the
European Union or the European Research Council. Neither the European
Union nor the granting authority can be held responsible for them.
Andrea Massa wishes to thank E. Vico for her never-ending inspiration,
support, guidance, and help.

~

\newpage
\section*{FIGURE CAPTIONS}

\begin{itemize}
\item \textbf{Figure 1.} Sketch of (\emph{a}) \emph{3D} geometry, (\emph{b})
the \emph{2D} top-view of the wireless communication scenario $\mathcal{D}$,
and (\emph{c}) the \emph{2D} geometry of the ideal scenario $\mathcal{D}'$.
\item \textbf{Figure 2.} \emph{Numerical Assessment} ($N=32$, $f=3.5$
{[}GHz{]}) - \emph{HFSS} model of the primary source $\Psi$.
\item \textbf{Figure 3.} \emph{Numerical Assessment} (\emph{{}``Orgiano''
Scenario} - $N=32$, $f=3.5$ {[}GHz{]}) - \emph{}Plots of (\emph{a})
top view of the outdoor scenario (\emph{OSM} Database) and (\emph{b})
the corresponding WinProp model.
\item \textbf{Figure 4.} \emph{Numerical Assessment} (\emph{{}``Orgiano''
Scenario} - $N=32$, $f=3.5$ {[}GHz{]}) - Magnitude of the simulated
(\emph{a})(\emph{d})(\emph{g}) $x$-component ($\left|E_{x}^{\left(n\right)}\left(x,\, y\right)\right|$),
(\emph{b})(\emph{e})(\emph{h}) $y$-component ($\left|E_{y}^{\left(n\right)}\left(x,\, y\right)\right|$),
and (\emph{c})(\emph{f})(\emph{i}) $z$-component ($\left|E_{z}^{\left(n\right)}\left(x,\, y\right)\right|$)
of the \emph{EPEP} of the $n$-th element of the \emph{BTS} array:
(\emph{a})-(\emph{c}) $n=1$, (\emph{d})-(\emph{f}) $n=12$, and (\emph{g})-(\emph{i})
$n=32$.
\item \textbf{Figure 5.} \emph{Numerical Assessment} (\emph{{}``Orgiano''
Scenario} - $N=32$, $f=3.5$ {[}GHz{]}) - Received power distribution
in $\mathcal{D}$ computed with (\emph{a})-(\emph{c}) the \emph{EPEP}
approach, $P_{rx}\left(\underline{r}\right)$, or (\emph{d})-(\emph{f})
by simulating the fully-excited \emph{BTS} antenna array, $\widehat{P}_{rx}\left(\underline{r}\right)$,
together with (\emph{g})-(\emph{i}) the corresponding absolute difference
maps, $\Delta\widehat{P}_{rx}\left(\underline{r}\right)$, when considering
(\emph{a})(\emph{d})(\emph{g}) uniform, (\emph{b})(\emph{e})(\emph{h})
steered, and (\emph{c})(\emph{f})(\emph{i}) random phase excitations.
\item \textbf{Figure 6.} \emph{Numerical Assessment} (\emph{{}``Orgiano''
Scenario} - $N=32$, $f=3.5$ {[}GHz{]}, $\zeta^{\max}=20$ {[}W{]})
- Maps of the received power distribution for (\emph{a}) the ideal
source $\Psi'$ radiating in free-space and the \emph{BTS} array radiating
within the complex-scattering scenario $\mathcal{D}$ when excited
with (\emph{b}) uniform or (\emph{c}) steered phase excitations.
\item \textbf{Figure 7.} \emph{Numerical Assessment} (\emph{{}``Orgiano''
Scenario} - $N=32$, $f=3.5$ {[}GHz{]}, $\zeta^{\max}=20$ {[}W{]})
- Behavior of the normalized cost function, $\hat{\Phi}\left(\underline{\beta}\right)$,
versus the iteration index, $i$. 
\item \textbf{Figure 8.} \emph{Numerical Assessment} (\emph{{}``Orgiano''
Scenario} - $N=32$, $f=3.5$ {[}GHz{]}, $\zeta^{\max}=20$ {[}W{]})
- Plots of (\emph{a}) the phase of the array excitations, $\underline{\beta}$,
and (\emph{b})-(\emph{d}) the normalized power patterns radiated by
$\Psi$ when setting (\emph{b}) $\underline{\beta}=\underline{\beta}^{opt}$,
(\emph{c}) $\underline{\beta}=\underline{\beta}^{uni}$, and (\emph{d})
$\underline{\beta}=\underline{\beta}^{ste}$.
\item \textbf{Figure 9.} \emph{Numerical Assessment} (\emph{{}``Orgiano''
Scenario} - $N=32$, $f=3.5$ {[}GHz{]}, $\zeta^{\max}=20$ {[}W{]})
- Plots of the optimized received power distribution, $P_{rx}^{opt}\left(\underline{r}\right)$,
in (\emph{a}) $\mathcal{D}$ and (\emph{b}) $\Omega$ along with (\emph{c})-(\emph{e})
the absolute difference map with respect to the target power distribution,
$\left|\Delta P_{rx}^{tar}\left(\underline{r}\right)\right|$, for
(\emph{c}) the \emph{OSS}, (\emph{d}) the {}``uniform'', and (\emph{e})
the {}``steered'' solutions.
\item \textbf{Figure 10.} \emph{Numerical Assessment} (\emph{{}``Orgiano''
Scenario} - $N=32$, $f=3.5$ {[}GHz{]}, $\zeta^{\max}=20$ {[}W{]})
- Map of the received power improvement, $\Delta P_{rx}^{opt}\left(\underline{r}\right)$,
yielded in $\Omega$ by the \emph{OSS} solution with respect to (\emph{a})
the {}``uniform'' and (\emph{b}) the {}``steered'' solutions.
\item \textbf{Figure 11.} \emph{Numerical Assessment} (\emph{{}``Padova''
Scenario} - $N=32$, $f=3.5$ {[}GHz{]}) - \emph{}Plots of (\emph{a})
the top view of the outdoor scenario (\emph{OSM} Database) and (\emph{b})
the corresponding WinProp model.
\item \textbf{Figure 12.} \emph{Numerical Assessment} (\emph{{}``Padova''
Scenario} - $N=32$, $f=3.5$ {[}GHz{]}, $\zeta^{\max}=20$ {[}W{]})
- Maps of the received power distribution for (\emph{a}) the ideal
source $\Psi'$ radiating in free-space and the \emph{BTS} array radiating
within the complex-scattering scenario $\mathcal{D}$ when excited
with (\emph{b}) uniform or (\emph{c}) steered phase excitations.
\item \textbf{Figure 13.} \emph{Numerical Assessment} (\emph{{}``Padova''
Scenario} - $N=32$, $f=3.5$ {[}GHz{]}, $\zeta^{\max}=20$ {[}W{]})
- Plots of (\emph{a}) the phase of the array excitations, $\underline{\beta}$,
and (\emph{b})(\emph{c}) the normalized power patterns radiated by
$\Psi$ when setting (\emph{b}) $\underline{\beta}=\underline{\beta}^{opt}$
and (\emph{c}) $\underline{\beta}=\underline{\beta}^{ste}$.
\item \textbf{Figure 14.} \emph{Numerical Assessment} (\emph{{}``Padova''
Scenario} - $N=32$, $f=3.5$ {[}GHz{]}, $\zeta^{\max}=20$ {[}W{]})
- Plots of the optimized received power distribution, $P_{rx}^{opt}\left(\underline{r}\right)$,
in (\emph{a}) $\mathcal{D}$ and (\emph{b}) $\Omega$ along with (\emph{c})-(\emph{e})
the absolute difference map with respect to the target power distribution,
$\left|\Delta P_{rx}^{tar}\left(\underline{r}\right)\right|$, for
(\emph{c}) the \emph{OSS}, (\emph{d}) the {}``uniform'', and (\emph{e})
the {}``steered'' solutions.
\item \textbf{Figure 15.} \emph{Numerical Assessment} (\emph{{}``Padova''
Scenario} - $N=32$, $f=3.5$ {[}GHz{]}, $\zeta^{\max}=20$ {[}W{]})
- Map of the received power improvement, $\Delta P_{rx}^{opt}\left(\underline{r}\right)$,
yielded in $\Omega$ by the \emph{OSS} solution with respect to (\emph{a})
the {}``uniform'' and (\emph{b}) the {}``steered'' solutions.
\item \textbf{Figure 16.} \emph{Numerical Assessment} (\emph{{}``Padova''
Scenario} - $N=32$, $f=3.5$ {[}GHz{]}, $\zeta^{\max}\in\left[20,\,500\right]$
{[}W{]}) - Plot of the cost function value at the convergence of the
\emph{OSS}, $\Phi\left(\underline{\beta}^{opt}\right)$, versus the
maximum radiated power by the \emph{BTS} array, $\zeta^{\max}$.
\item \textbf{Figure 17.} \emph{Numerical Assessment} (\emph{{}``Padova''
Scenario} - $N=32$, $f=3.5$ {[}GHz{]}, $\zeta^{\max}\in\left[20,\,500\right]$
{[}W{]}) - \emph{OSS} normalized power pattern when (\emph{a}) $\delta=2$
($\rightarrow\zeta^{\max}=80$ {[}W{]}), (\emph{b}) $\delta=3$ ($\rightarrow\zeta^{\max}=180$
{[}W{]}), (\emph{c}) $\delta=4$ ($\rightarrow\zeta^{\max}=320$ {[}W{]}),
and (\emph{d}) $\delta=5$ ($\rightarrow\zeta^{\max}=500$ {[}W{]}).
\item \textbf{Figure 18.} \emph{Numerical Assessment} (\emph{{}``Padova''
Scenario} - $N=32$, $f=3.5$ {[}GHz{]}, $\zeta^{\max}\in\left[20,\,500\right]$
{[}W{]}) - \emph{}Optimized received power distribution, $P_{rx}^{opt}\left(\underline{r}\right)$,
$\underline{r}\in\mathcal{D}$, when (\emph{a}) $\delta=2$ ($\rightarrow\zeta^{\max}=80$
{[}W{]}), (\emph{b}) $\delta=3$ ($\rightarrow\zeta^{\max}=180$ {[}W{]}),
(\emph{c}) $\delta=4$ ($\rightarrow\zeta^{\max}=320$ {[}W{]}), and
(\emph{d}) $\delta=5$ ($\rightarrow\zeta^{\max}=500$ {[}W{]}).
\item \textbf{Figure 19.} \emph{Numerical Assessment} (\emph{{}``Padova''
Scenario} - $N=32$, $f=3.5$ {[}GHz{]}, $\zeta^{\max}\in\left[20,\,500\right]$
{[}W{]}) - \emph{}Maps of the target distribution mismatch, $\left|\Delta P_{rx}^{tar}\left(\underline{r}\right)\right|$,
$\underline{r}\in\Omega$, when (\emph{a}) $\delta=2$ ($\rightarrow\zeta^{\max}=80$
{[}W{]}), (\emph{b}) $\delta=3$ ($\rightarrow\zeta^{\max}=180$ {[}W{]}),
(\emph{c}) $\delta=4$ ($\rightarrow\zeta^{\max}=320$ {[}W{]}), and
(\emph{d}) $\delta=5$ ($\rightarrow\zeta^{\max}=500$ {[}W{]}).
\item \textbf{Figure 20.} \emph{Numerical Assessment} (\emph{{}``Padova''
Scenario} - $N=32$, $f=3.5$ {[}GHz{]}, $\zeta^{\max}=20$ {[}W{]})
- Plots of (\emph{a})(\emph{b}) the \emph{OSS} normalized power pattern,
(\emph{c})(\emph{d}) the corresponding received power distribution
$P_{rx}^{opt}\left(\underline{r}\right)$ $\underline{r}\in\mathcal{D}$,
and (\emph{e})(\emph{f}) map of the target distribution mismatch,
$\left|\Delta P_{rx}^{tar}\left(\underline{r}\right)\right|$, when
opportunistically exploiting the environment to enhance the coverage
within the \emph{RoI}s (\emph{a})(\emph{c})(\emph{e}) $\Omega_{2}$
or (\emph{b})(\emph{d})(\emph{f}) $\Omega_{3}$.
\end{itemize}

\section*{TABLE CAPTIONS}

\begin{itemize}
\item \textbf{Table I.} \emph{Numerical Assessment} ($N=32$, $f=3.5$ {[}GHz{]},
$\zeta^{\max}=20$ {[}W{]}) - Received power statistics.
\item \textbf{Table II.} \emph{Numerical Assessment} (\emph{{}``Padova''
Scenario} - $N=32$, $f=3.5$ {[}GHz{]}) - \emph{}Received power statistics.
\end{itemize}
\newpage
\begin{center}~\vfill\end{center}

\begin{center}\begin{tabular}{c}
\includegraphics[%
  width=0.50\columnwidth]{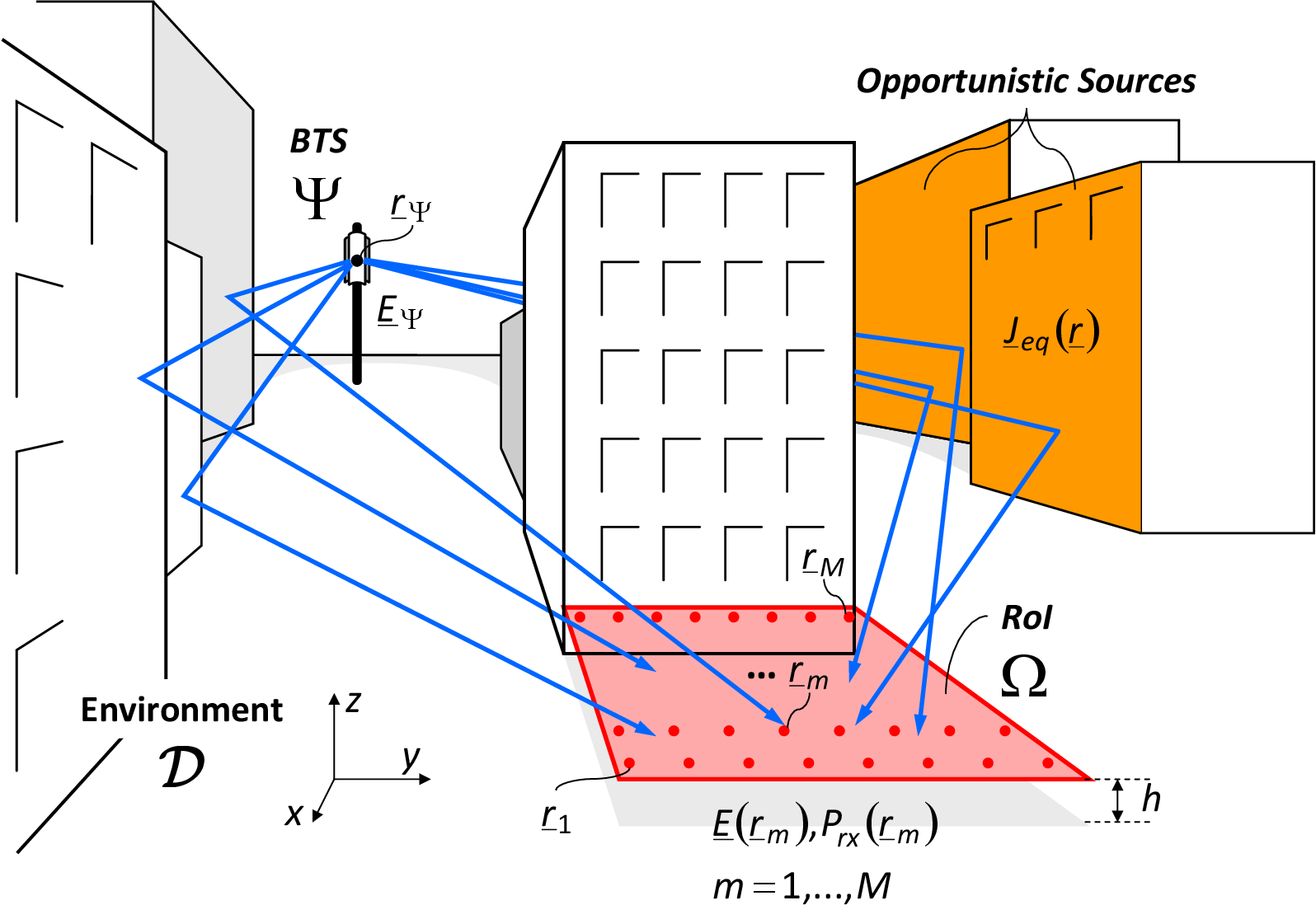}\tabularnewline
(\emph{a})\tabularnewline
\tabularnewline
\includegraphics[%
  width=0.50\columnwidth]{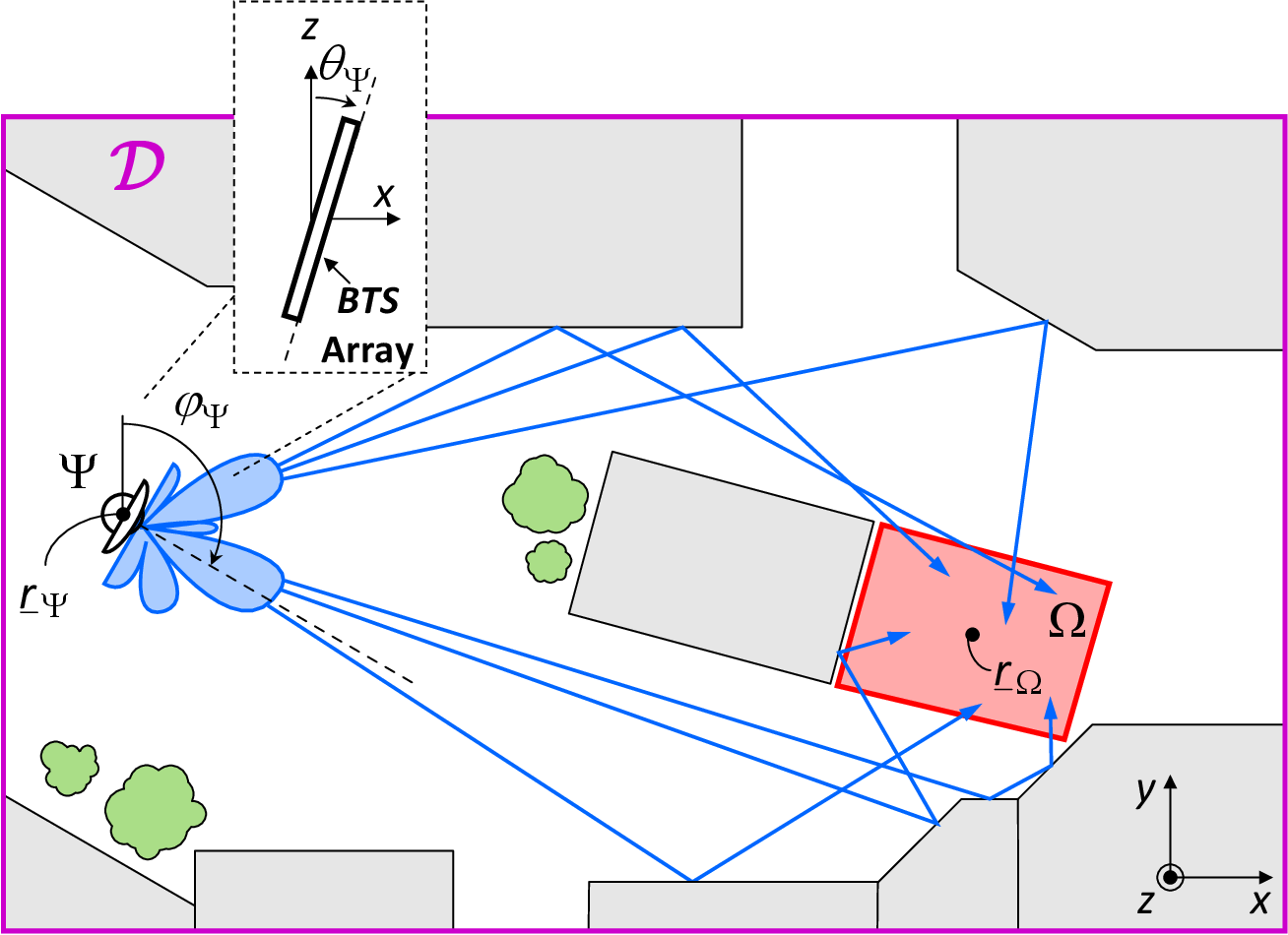}\tabularnewline
(\emph{b})\tabularnewline
\tabularnewline
\includegraphics[%
  width=0.50\columnwidth]{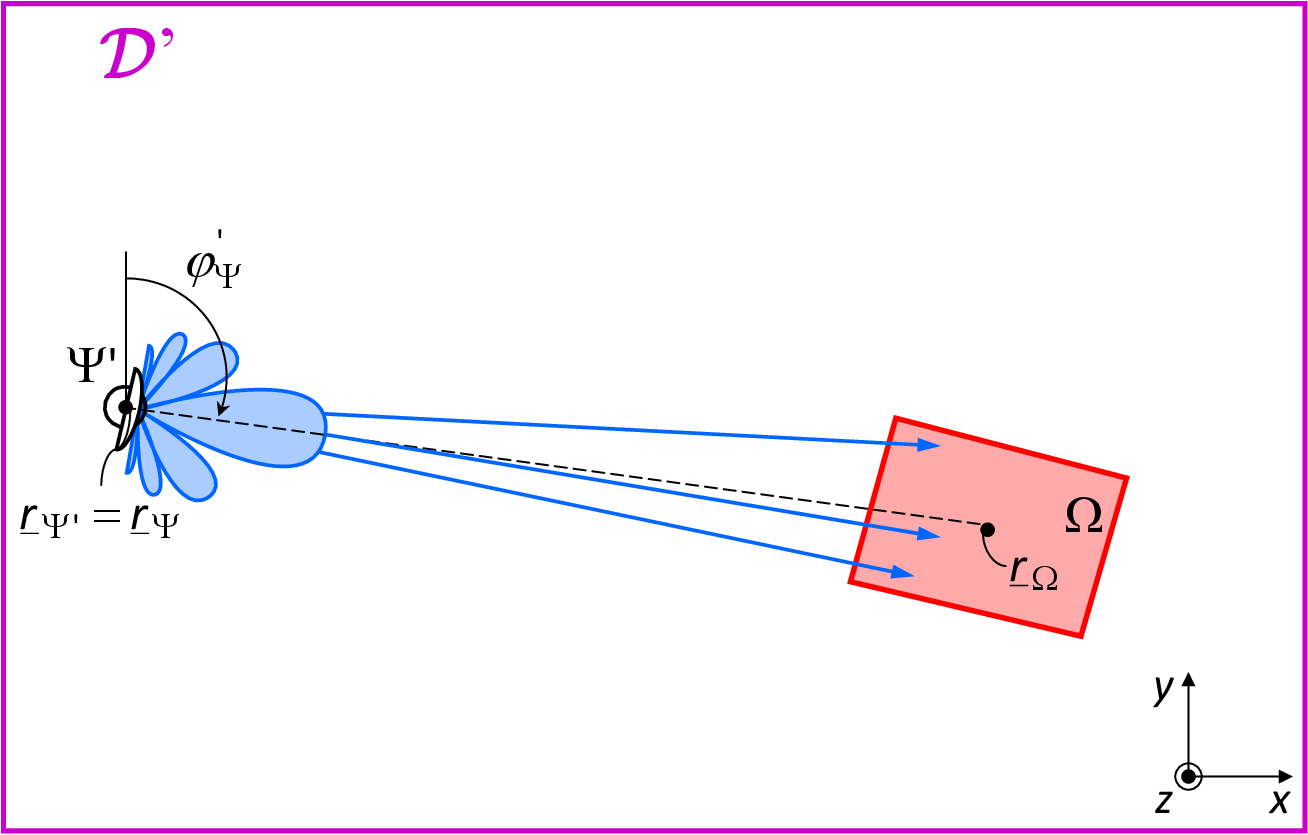}\tabularnewline
(\emph{c})\tabularnewline
\end{tabular}\end{center}

\begin{center}~\vfill\end{center}

\begin{center}\textbf{Fig. 1 - P. Da Ru} \textbf{\emph{et al.}}, {}``An
Opportunistic Source Synthesis Method ...''\end{center}

\newpage
\begin{center}~\vfill\end{center}

\begin{center}\includegraphics[%
  width=0.80\columnwidth,
  keepaspectratio]{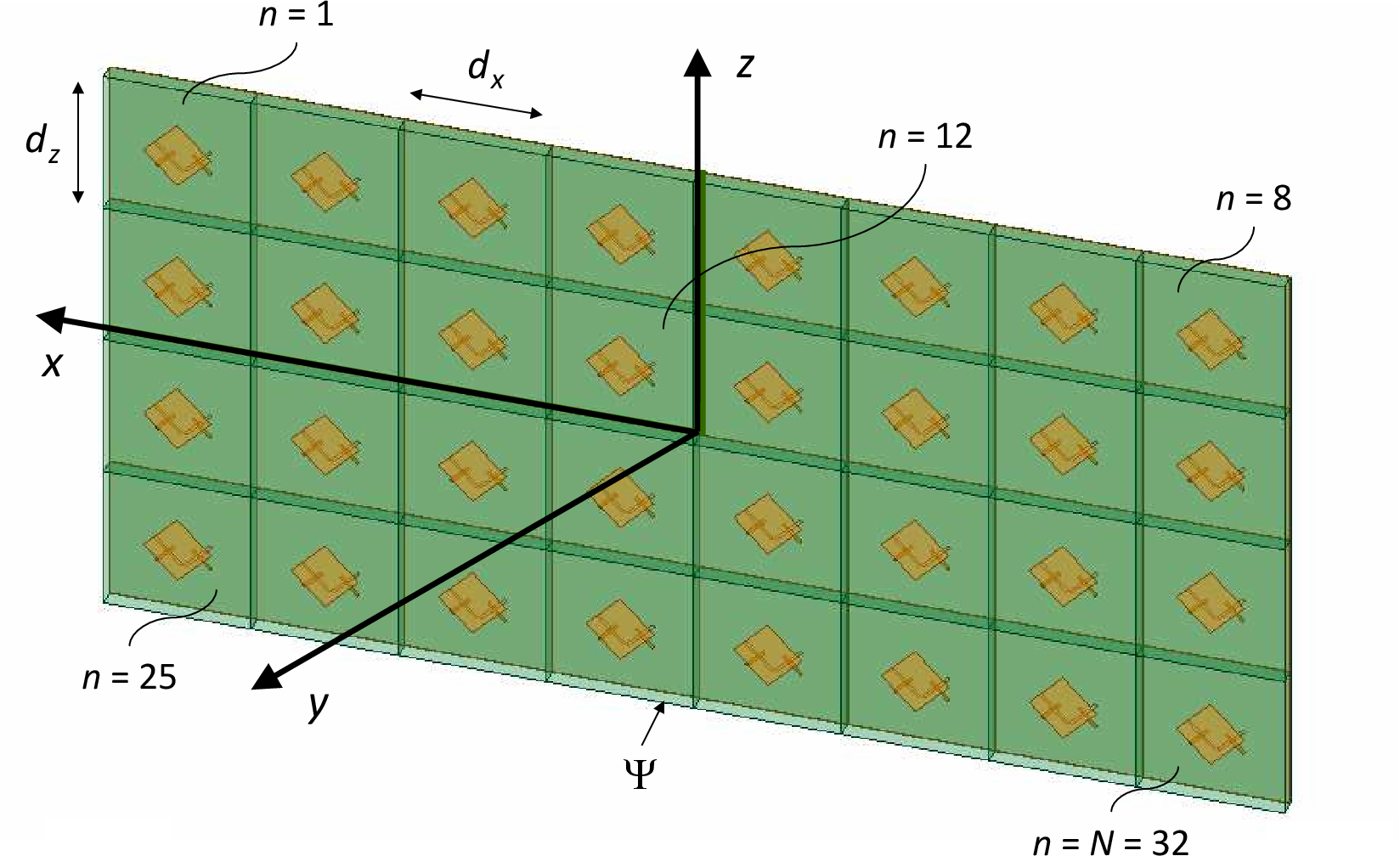}\end{center}

\begin{center}~\vfill\end{center}

\begin{center}\textbf{Fig. 2 - P. Da Ru} \textbf{\emph{et al.}}, {}``An
Opportunistic Source Synthesis Method ...''\end{center}

\newpage
\begin{center}~\vfill\end{center}

\begin{center}\begin{tabular}{c}
\includegraphics[%
  width=0.40\columnwidth]{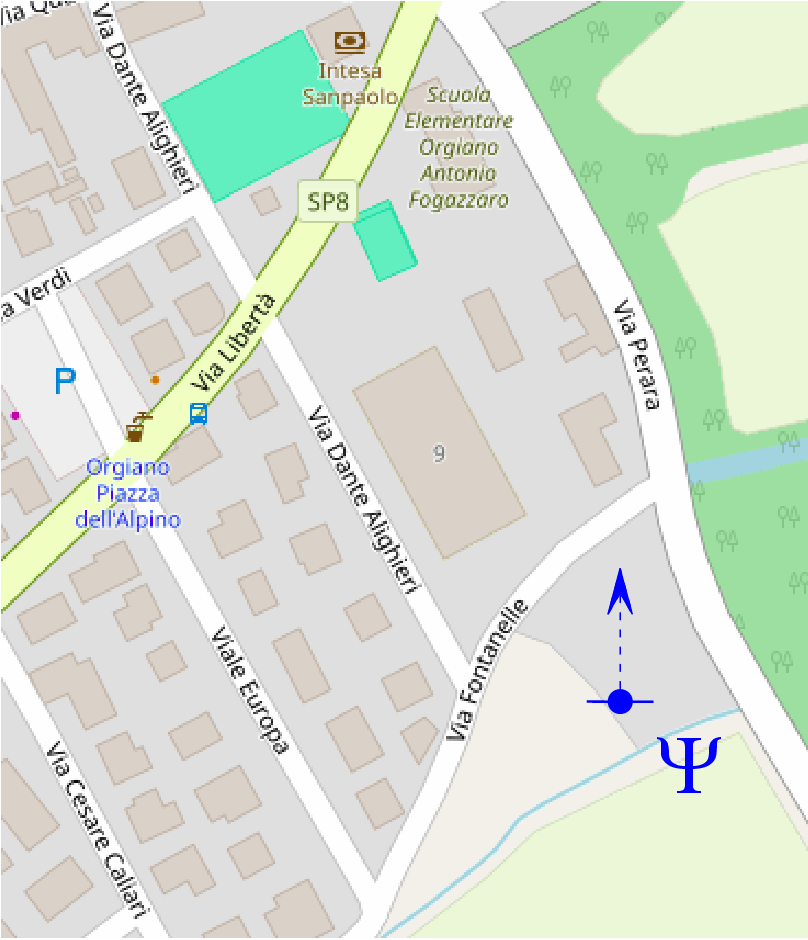}\tabularnewline
(\emph{a})\tabularnewline
\tabularnewline
\includegraphics[%
  width=0.45\columnwidth]{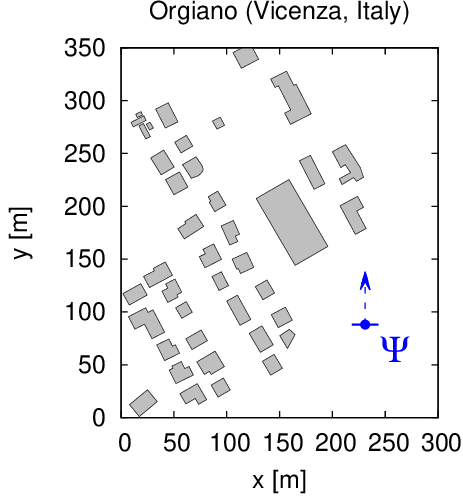}\tabularnewline
(\emph{b})\tabularnewline
\end{tabular}\end{center}

\begin{center}~\vfill\end{center}

\begin{center}\textbf{Fig. 3 - P. Da Ru} \textbf{\emph{et al.}}, {}``An
Opportunistic Source Synthesis Method ...''\end{center}

\newpage
\begin{center}~\vfill\end{center}

\begin{center}\begin{tabular}{cccc}
\begin{sideways}
\end{sideways}&
$\left|E_{x}^{\left(n\right)}\right|$&
$\left|E_{y}^{\left(n\right)}\right|$&
$\left|E_{z}^{\left(n\right)}\right|$\tabularnewline
\begin{sideways}
~~~~~~~~~~~~~~~~~$n=1$%
\end{sideways}&
\includegraphics[%
  width=0.33\columnwidth]{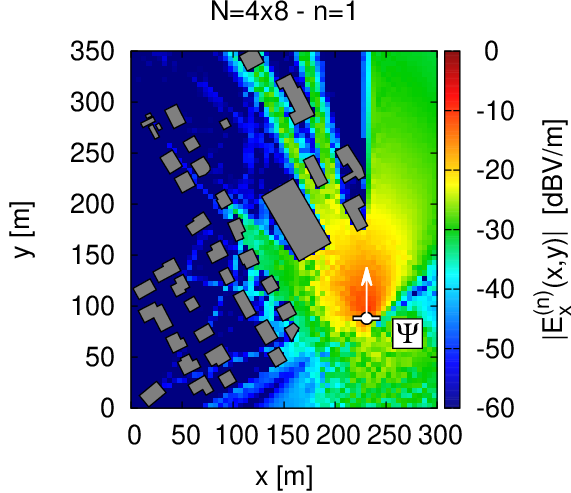}&
\includegraphics[%
  width=0.33\columnwidth]{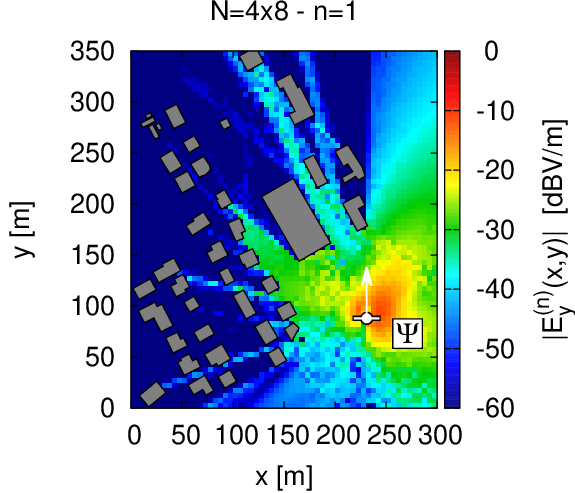}&
\includegraphics[%
  width=0.33\columnwidth]{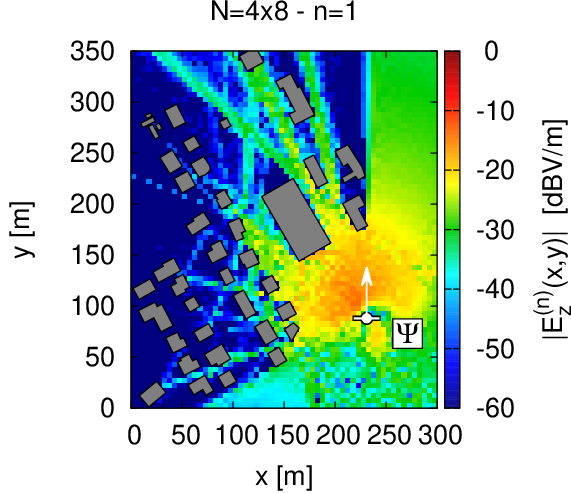}\tabularnewline
\begin{sideways}
\end{sideways}&
(\emph{a})&
(\emph{b})&
(\emph{c})\tabularnewline
\begin{sideways}
\end{sideways}&
&
&
\tabularnewline
\begin{sideways}
~~~~~~~~~~~~~~~~~$n=12$%
\end{sideways}&
\includegraphics[%
  width=0.33\columnwidth]{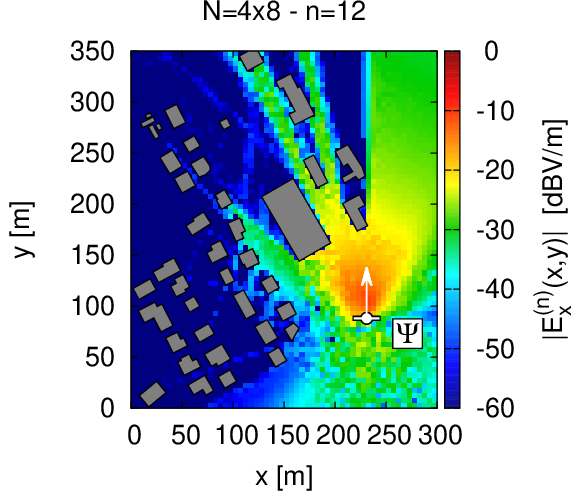}&
\includegraphics[%
  width=0.33\columnwidth]{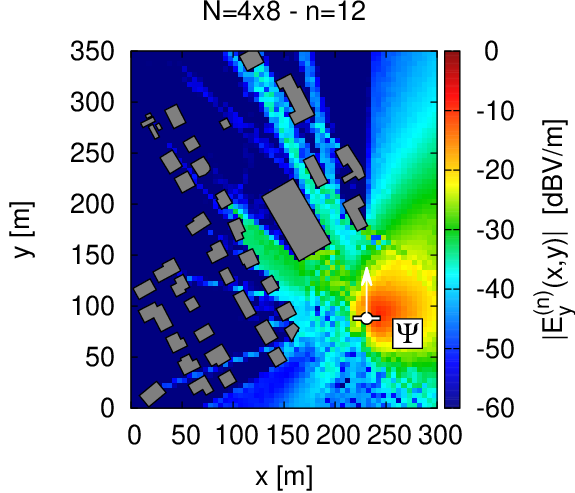}&
\includegraphics[%
  width=0.33\columnwidth]{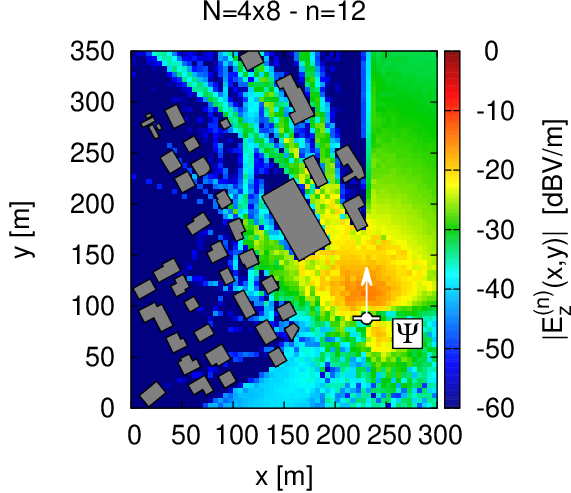}\tabularnewline
\begin{sideways}
\end{sideways}&
(\emph{d})&
(\emph{e})&
(\emph{f})\tabularnewline
\begin{sideways}
\end{sideways}&
&
&
\tabularnewline
\begin{sideways}
~~~~~~~~~~~~~~~~~$n=32$%
\end{sideways}&
\includegraphics[%
  width=0.33\columnwidth]{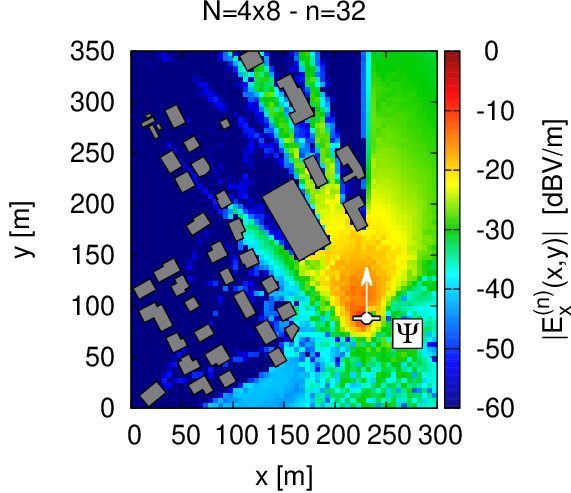}&
\includegraphics[%
  width=0.33\columnwidth]{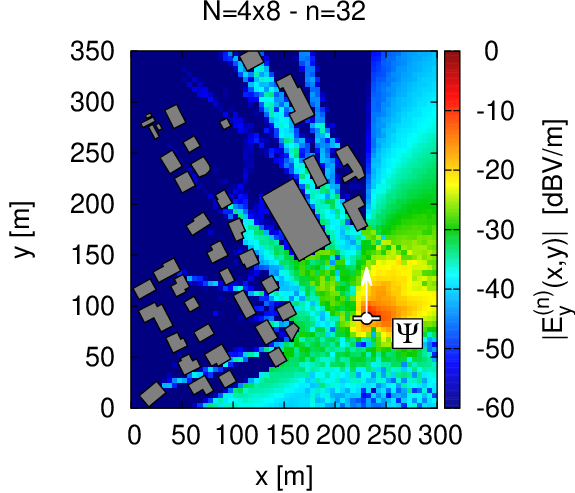}&
\includegraphics[%
  width=0.33\columnwidth]{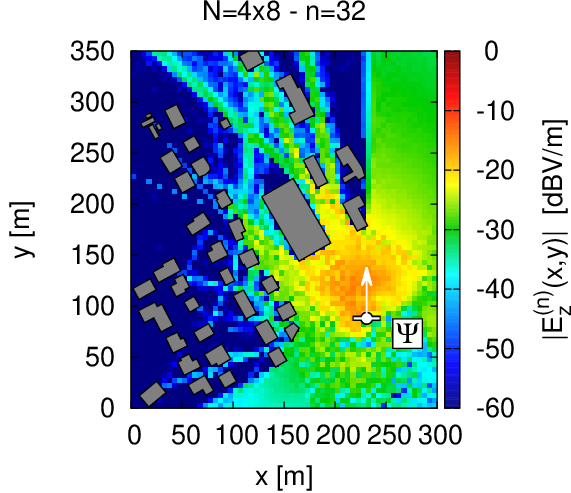}\tabularnewline
\begin{sideways}
\end{sideways}&
(\emph{g})&
(\emph{h})&
(\emph{i})\tabularnewline
\end{tabular}\end{center}

\begin{center}~\vfill\end{center}

\begin{center}\textbf{Fig. 4 - P. Da Ru} \textbf{\emph{et al.}}, {}``An
Opportunistic Source Synthesis Method ...''\end{center}

\newpage
\begin{center}~\vfill\end{center}

\begin{center}\begin{tabular}{ccc}
\includegraphics[%
  width=0.33\columnwidth]{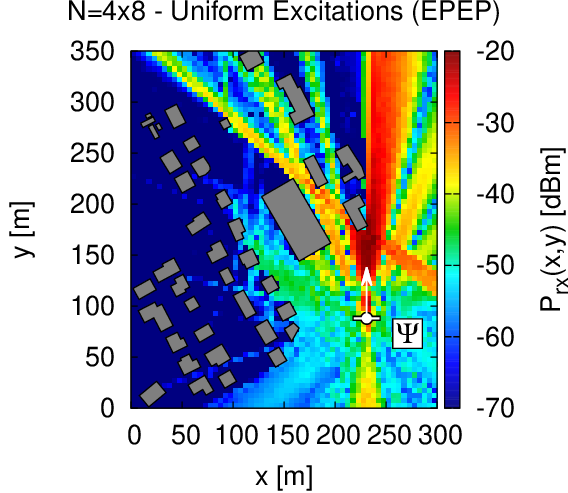}&
\includegraphics[%
  width=0.33\columnwidth]{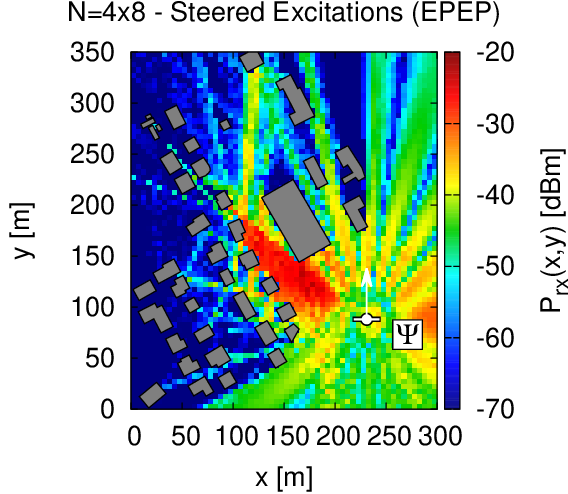}&
\includegraphics[%
  width=0.33\columnwidth]{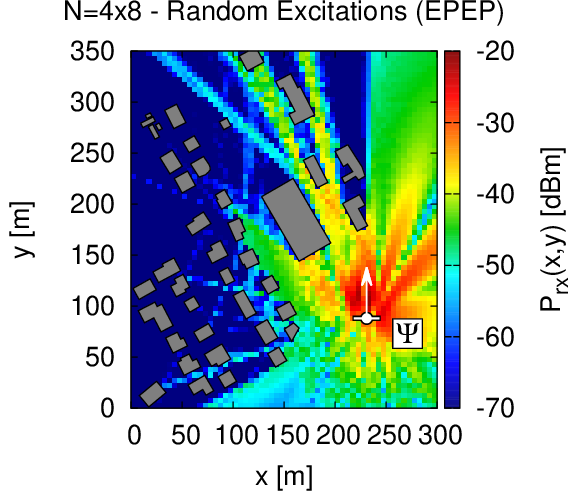}\tabularnewline
(\emph{a})&
(\emph{b})&
(\emph{c})\tabularnewline
&
&
\tabularnewline
\includegraphics[%
  width=0.33\columnwidth]{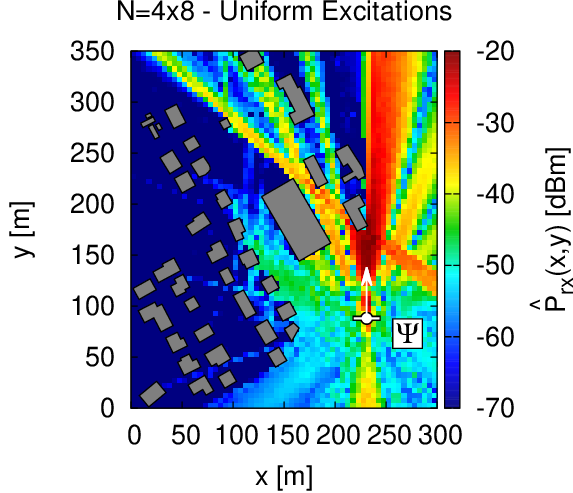}&
\includegraphics[%
  width=0.33\columnwidth]{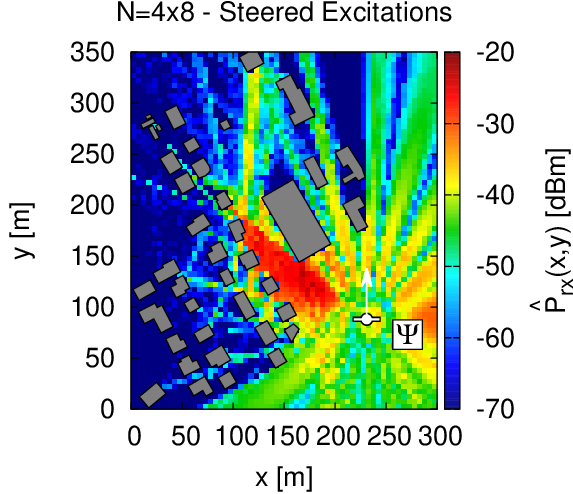}&
\includegraphics[%
  width=0.33\columnwidth]{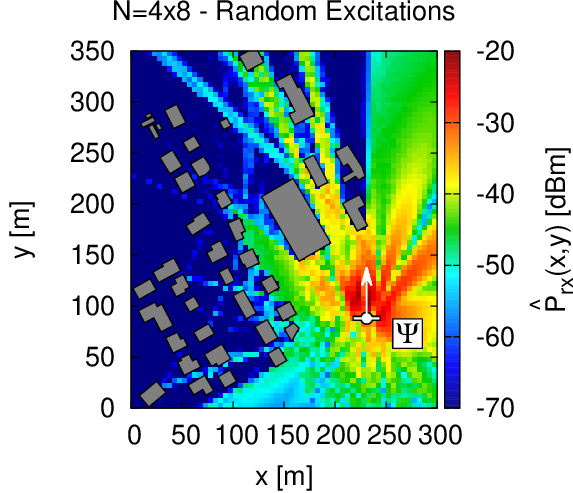}\tabularnewline
(\emph{d})&
(\emph{e})&
(\emph{f})\tabularnewline
&
&
\tabularnewline
\includegraphics[%
  width=0.33\columnwidth]{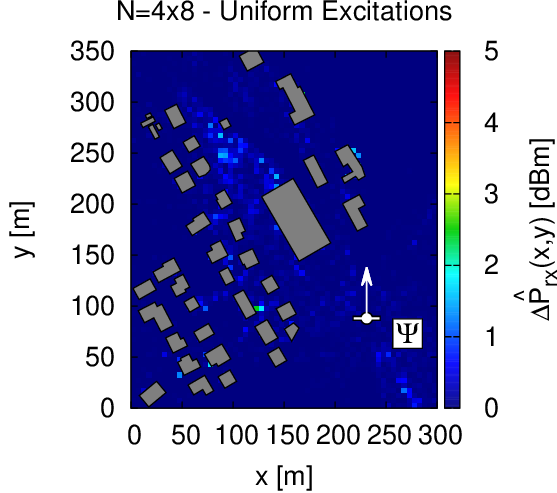}&
\includegraphics[%
  width=0.33\columnwidth]{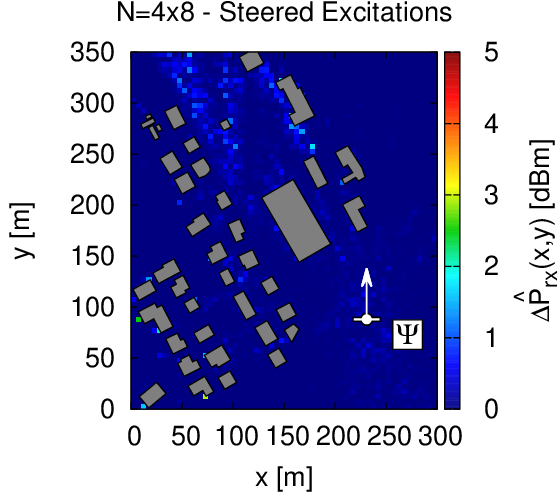}&
\includegraphics[%
  width=0.33\columnwidth]{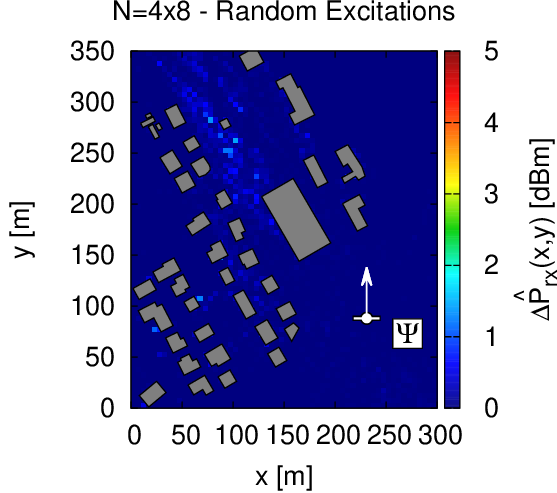}\tabularnewline
(\emph{g})&
(\emph{h})&
(\emph{i})\tabularnewline
\end{tabular}\end{center}

\begin{center}~\vfill\end{center}

\begin{center}\textbf{Fig. 5 - P. Da Ru} \textbf{\emph{et al.}}, {}``An
Opportunistic Source Synthesis Method ...''\end{center}

\newpage
\begin{center}~\vfill\end{center}

\begin{center}\begin{tabular}{c}
\includegraphics[%
  width=0.40\columnwidth]{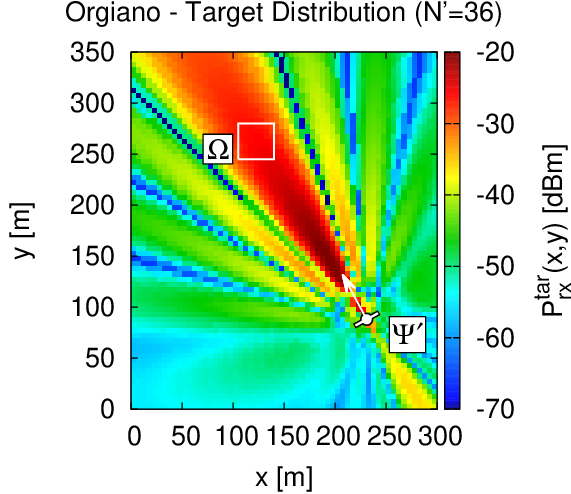}\tabularnewline
(\emph{a})\tabularnewline
\tabularnewline
\includegraphics[%
  width=0.40\columnwidth]{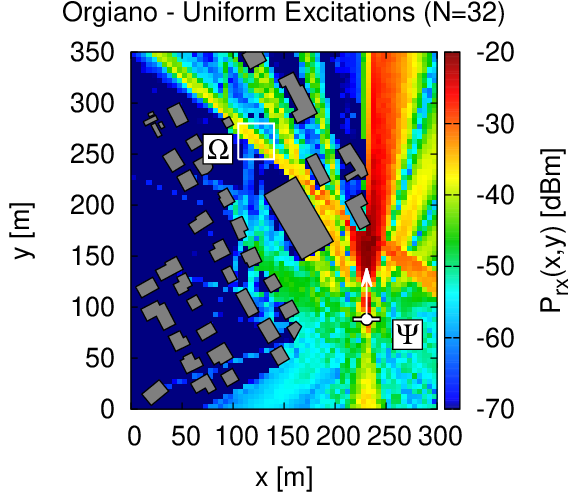}\tabularnewline
(\emph{b})\tabularnewline
\tabularnewline
\includegraphics[%
  width=0.40\columnwidth]{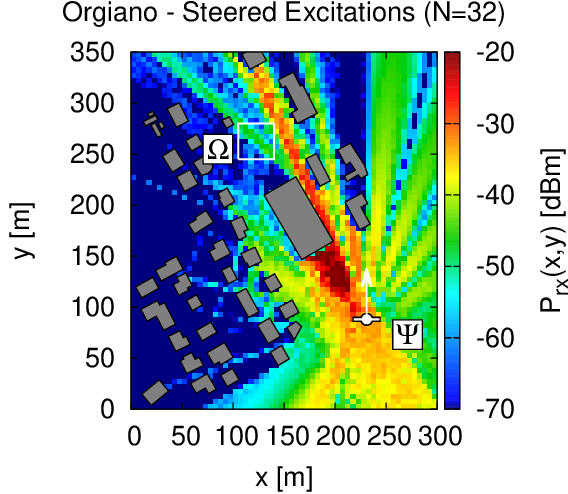}\tabularnewline
(\emph{c})\tabularnewline
\end{tabular}\end{center}

\begin{center}~\vfill\end{center}

\begin{center}\textbf{Fig. 6 - P. Da Ru} \textbf{\emph{et al.}}, {}``An
Opportunistic Source Synthesis Method ...''\end{center}

\newpage
\begin{center}~\vfill\end{center}

\begin{center}\includegraphics[%
  width=0.80\columnwidth]{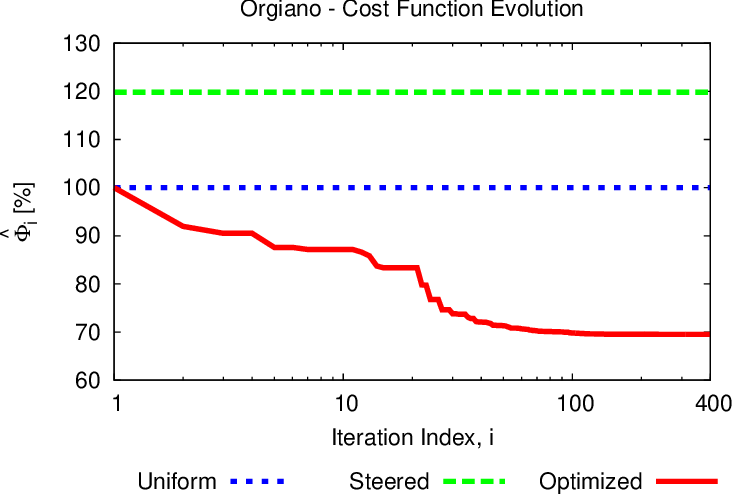}\end{center}

\begin{center}~\vfill\end{center}

\begin{center}\textbf{Fig. 7 - P. Da Ru} \textbf{\emph{et al.}}, {}``An
Opportunistic Source Synthesis Method ...''\end{center}

\newpage
\begin{center}~\vfill\end{center}

\begin{center}\begin{tabular}{cc}
\multicolumn{2}{c}{\includegraphics[%
  width=0.50\columnwidth]{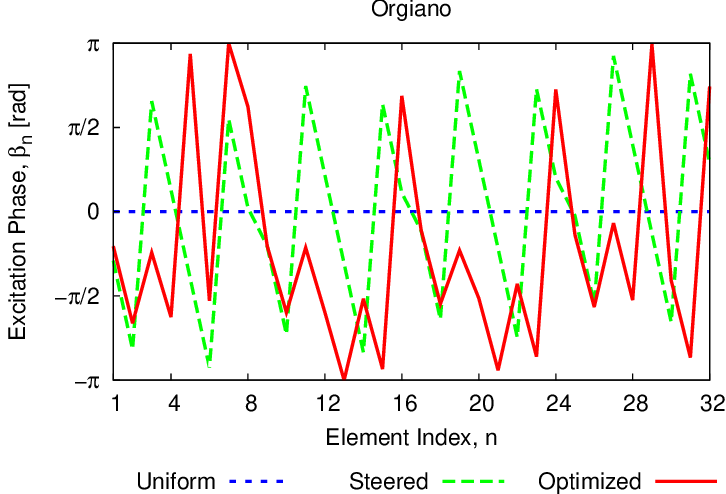}}\tabularnewline
\multicolumn{2}{c}{(\emph{a})}\tabularnewline
\multicolumn{2}{c}{}\tabularnewline
\multicolumn{2}{c}{\includegraphics[%
  width=0.40\columnwidth]{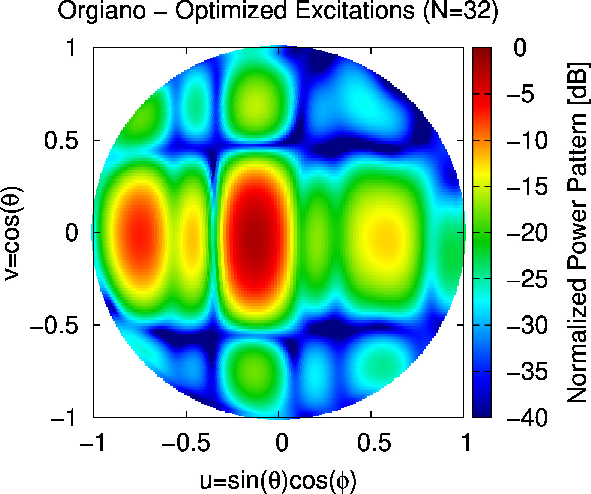}}\tabularnewline
\multicolumn{2}{c}{(\emph{b})}\tabularnewline
\multicolumn{2}{c}{}\tabularnewline
\includegraphics[%
  width=0.40\columnwidth]{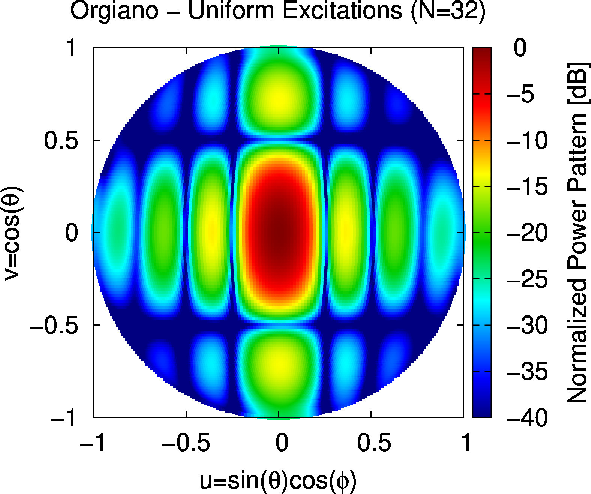}&
\includegraphics[%
  width=0.40\columnwidth]{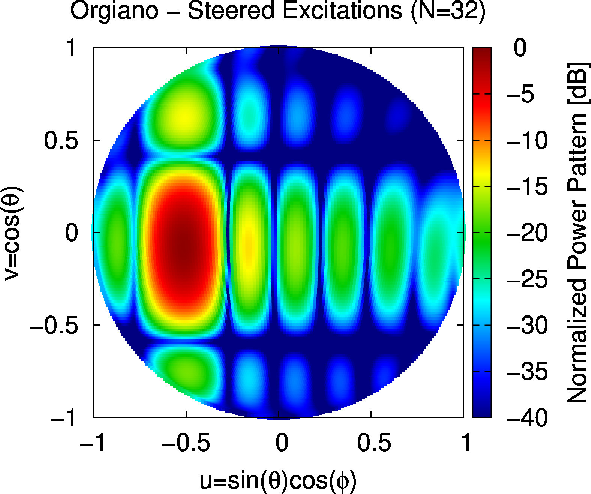}\tabularnewline
(\emph{c})&
(\emph{d})\tabularnewline
\end{tabular}\end{center}

\begin{center}~\vfill\end{center}

\begin{center}\textbf{Fig. 8 - P. Da Ru} \textbf{\emph{et al.}}, {}``An
Opportunistic Source Synthesis Method ...''\end{center}

\newpage
\begin{center}~\vfill\end{center}

\begin{center}\begin{tabular}{cc}
\multicolumn{2}{c}{\includegraphics[%
  width=0.50\columnwidth]{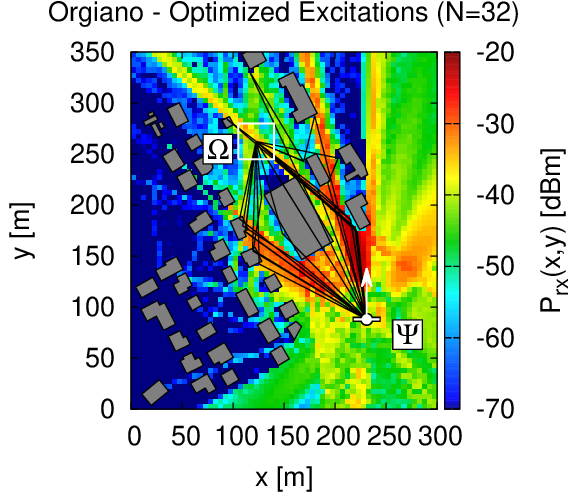}}\tabularnewline
\multicolumn{2}{c}{(\emph{a})}\tabularnewline
\multicolumn{2}{c}{}\tabularnewline
\includegraphics[%
  width=0.40\columnwidth]{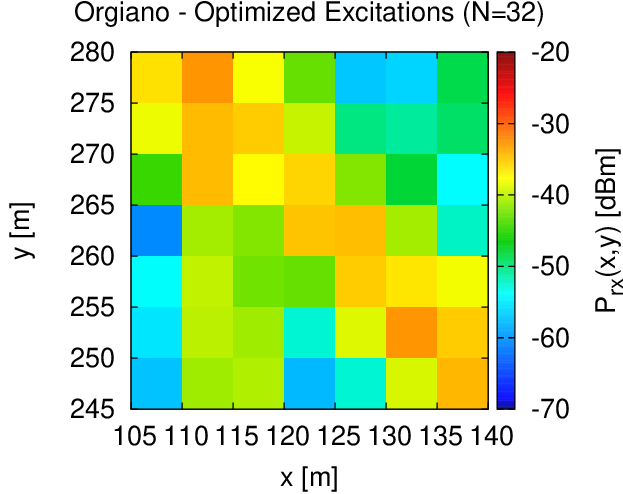}&
\includegraphics[%
  width=0.40\columnwidth]{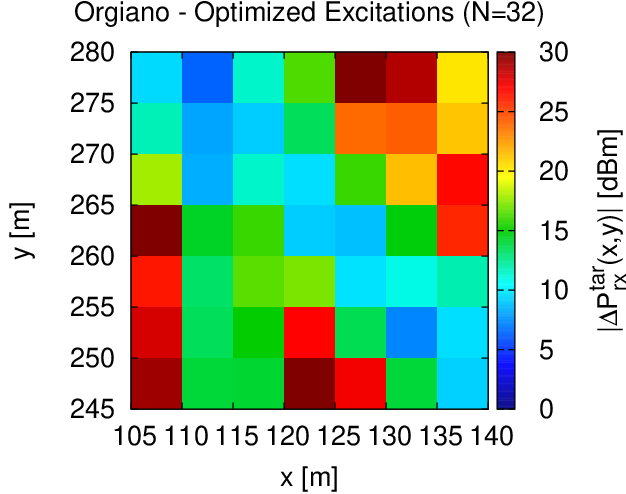} \tabularnewline
(\emph{b})&
(\emph{c})\tabularnewline
&
\tabularnewline
\includegraphics[%
  width=0.40\columnwidth]{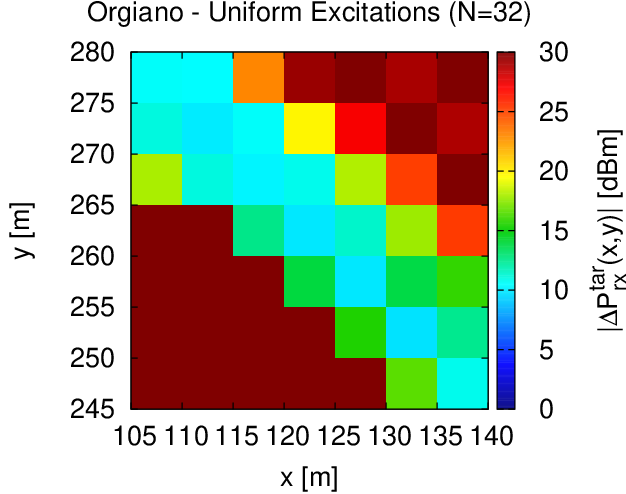} &
\includegraphics[%
  width=0.40\columnwidth]{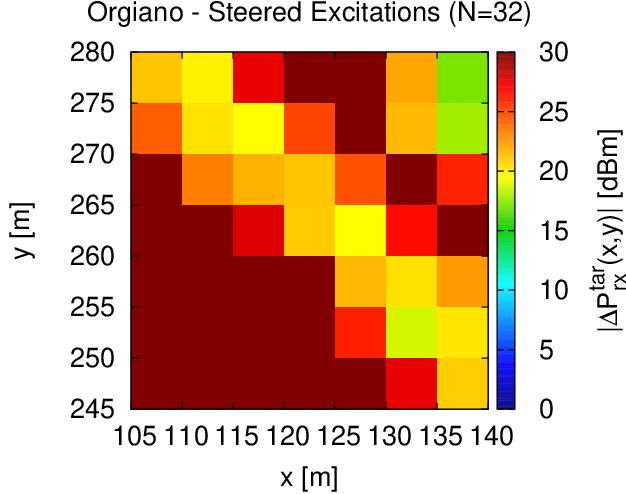} \tabularnewline
(\emph{d})&
(\emph{e})\tabularnewline
\end{tabular}\end{center}

\begin{center}\textbf{Fig. 9 - P. Da Ru} \textbf{\emph{et al.}}, {}``An
Opportunistic Source Synthesis Method ...''\end{center}

\newpage
\begin{center}~\vfill\end{center}

\begin{center}\begin{tabular}{c}
\includegraphics[%
  width=0.50\columnwidth]{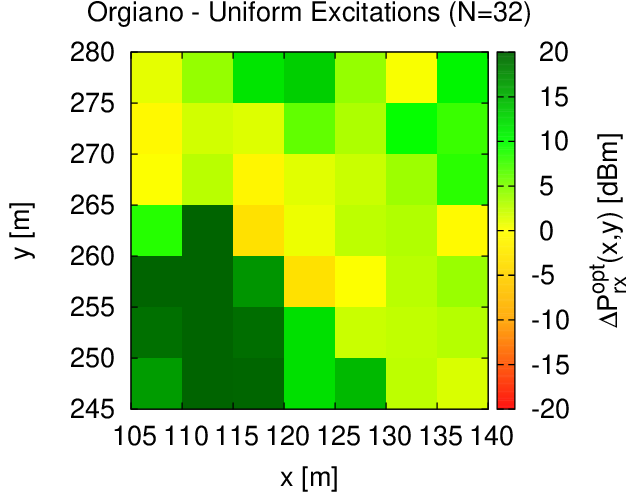}\tabularnewline
(\emph{a})\tabularnewline
\tabularnewline
\includegraphics[%
  width=0.50\columnwidth]{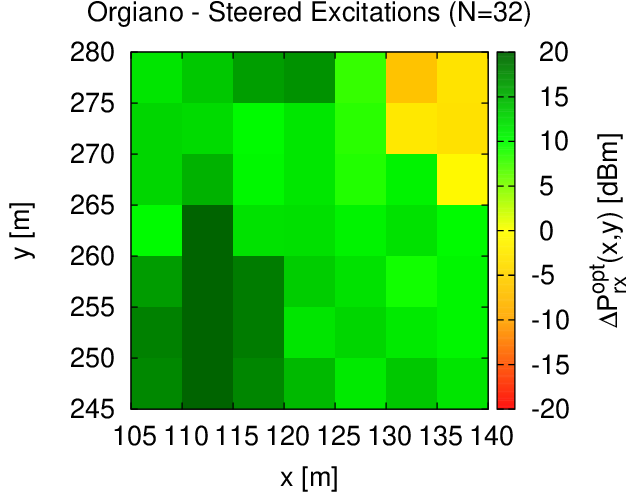}\tabularnewline
(\emph{b})\tabularnewline
\end{tabular}\end{center}

\begin{center}~\vfill\end{center}

\begin{center}\textbf{Fig. 10 - P. Da Ru} \textbf{\emph{et al.}},
{}``An Opportunistic Source Synthesis Method ...''\end{center}

\newpage
\noindent \begin{center}~\vfill\end{center}

\begin{center}\begin{tabular}{c}
\includegraphics[%
  width=0.60\columnwidth]{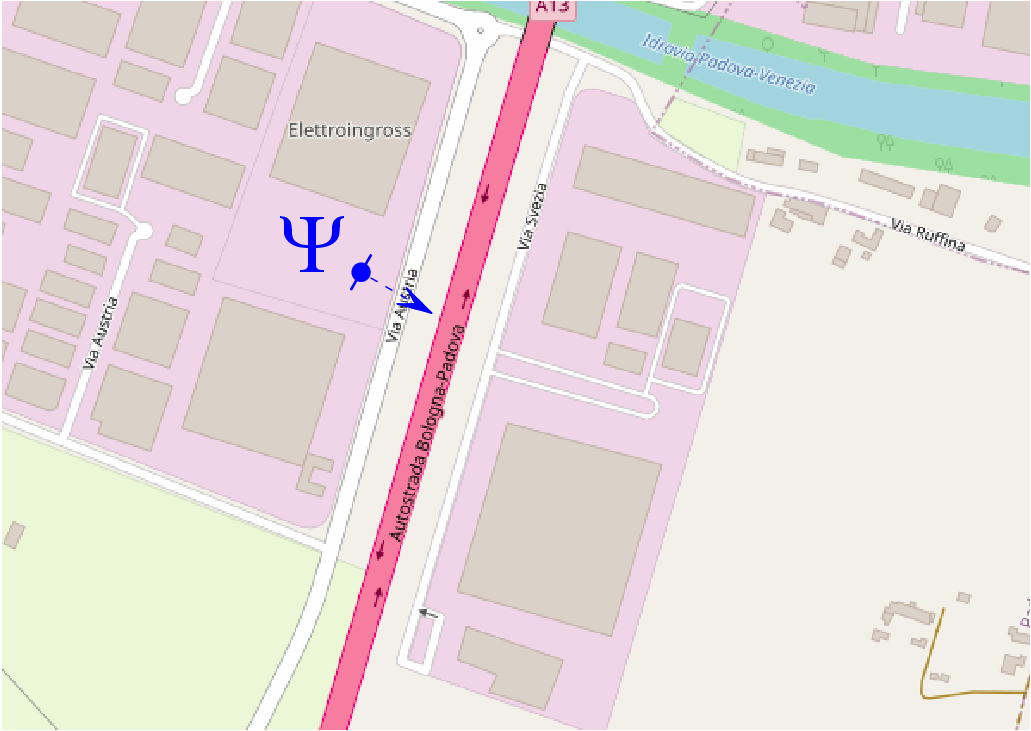}\tabularnewline
(\emph{a})\tabularnewline
\tabularnewline
\includegraphics[%
  width=0.60\columnwidth]{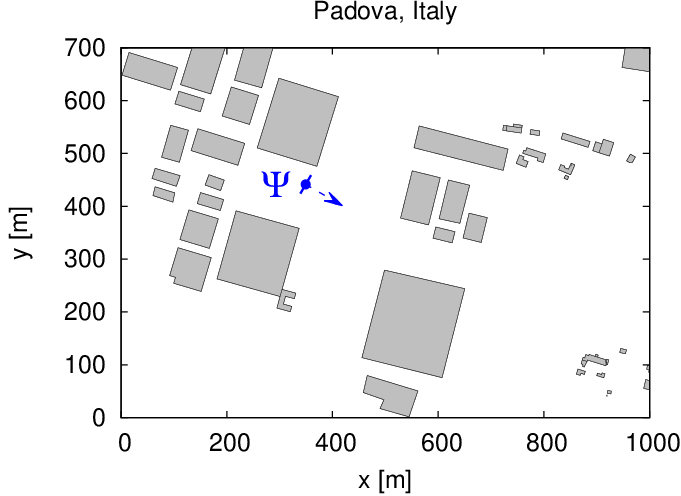}\tabularnewline
(\emph{b})\tabularnewline
\end{tabular}\end{center}

\begin{center}~\vfill\end{center}

\begin{center}\textbf{Fig. 11 - P. Da Ru} \textbf{\emph{et al.}},
{}``An Opportunistic Source Synthesis Method ...''\end{center}

\newpage
\begin{center}~\vfill\end{center}

\begin{center}\begin{tabular}{c}
\includegraphics[%
  width=0.50\columnwidth]{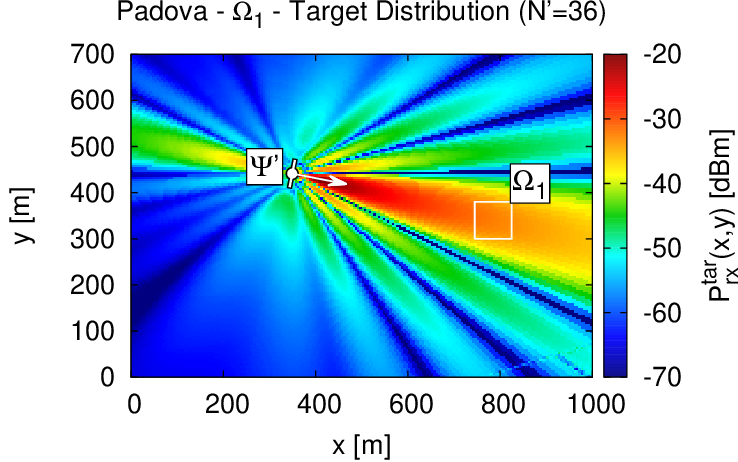}\tabularnewline
(\emph{a})\tabularnewline
\tabularnewline
\includegraphics[%
  width=0.50\columnwidth]{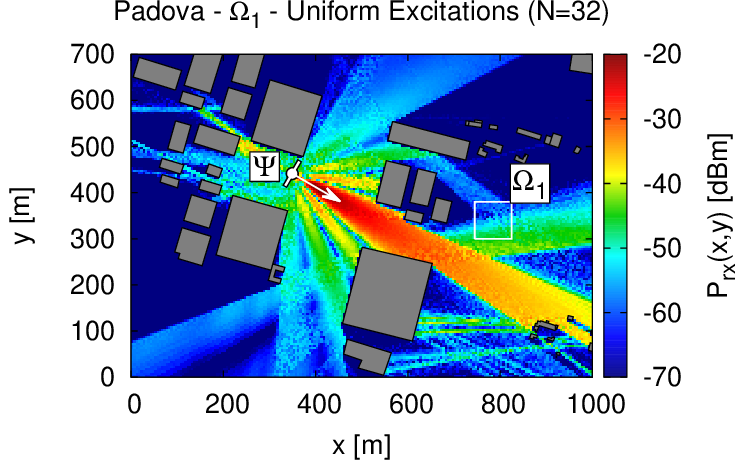}\tabularnewline
(\emph{b})\tabularnewline
\tabularnewline
\includegraphics[%
  width=0.50\columnwidth]{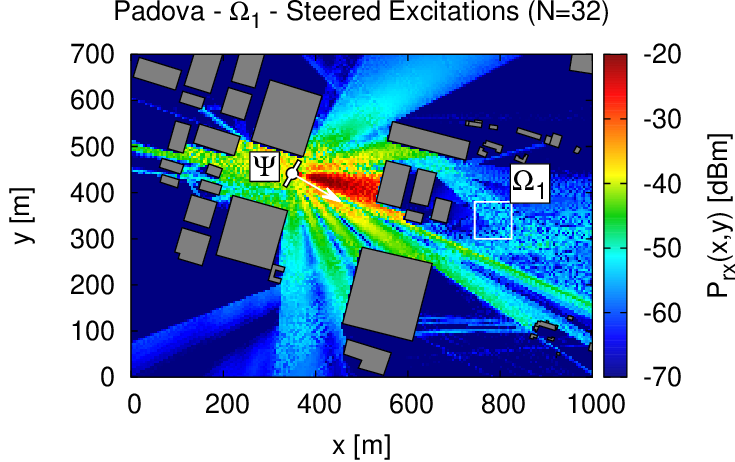}\tabularnewline
(\emph{c})\tabularnewline
\end{tabular}\end{center}

\begin{center}~\vfill\end{center}

\begin{center}\textbf{Fig. 12 - P. Da Ru} \textbf{\emph{et al.}},
{}``An Opportunistic Source Synthesis Method ...''\end{center}

\newpage
\begin{center}~\vfill\end{center}

\begin{center}\begin{tabular}{cc}
\multicolumn{2}{c}{\includegraphics[%
  width=0.60\columnwidth]{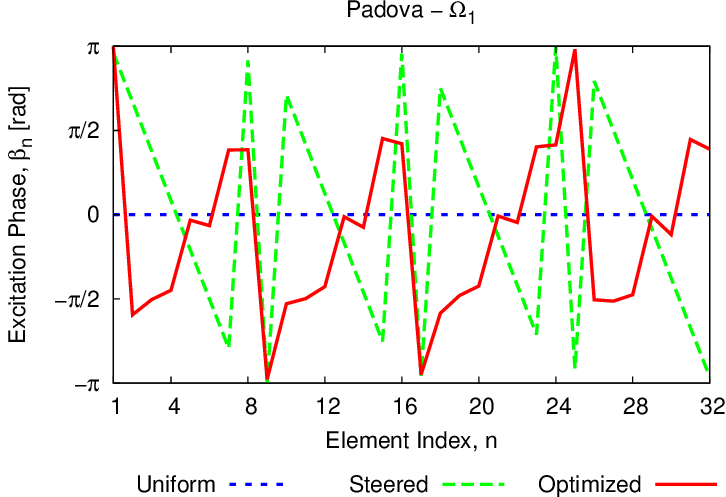}}\tabularnewline
\multicolumn{2}{c}{(\emph{a})}\tabularnewline
\multicolumn{2}{c}{}\tabularnewline
\includegraphics[%
  width=0.45\columnwidth]{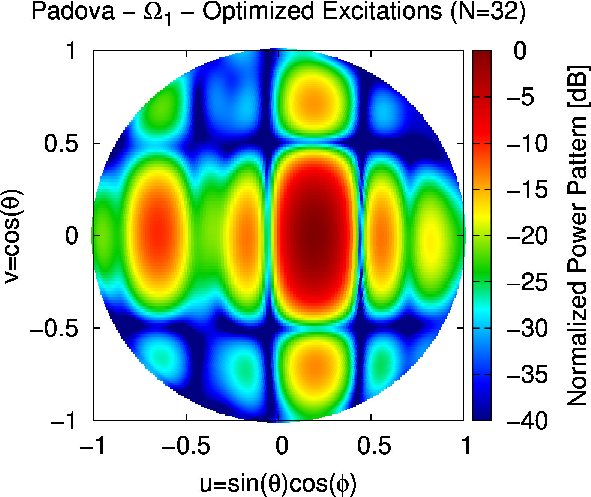}&
\includegraphics[%
  width=0.45\columnwidth]{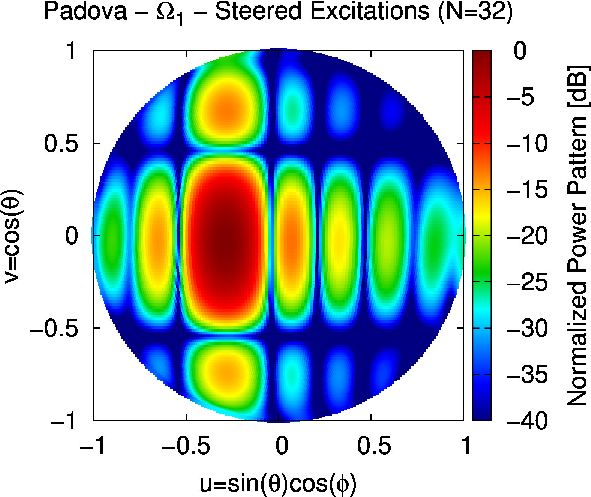}\tabularnewline
(\emph{b})&
(\emph{c})\tabularnewline
\end{tabular}\end{center}

\begin{center}~\vfill\end{center}

\begin{center}\textbf{Fig. 13 - P. Da Ru} \textbf{\emph{et al.}},
{}``An Opportunistic Source Synthesis Method ...''\end{center}

\newpage
\begin{center}~\vfill\end{center}

\begin{center}\begin{tabular}{cc}
\multicolumn{2}{c}{\includegraphics[%
  width=0.50\columnwidth]{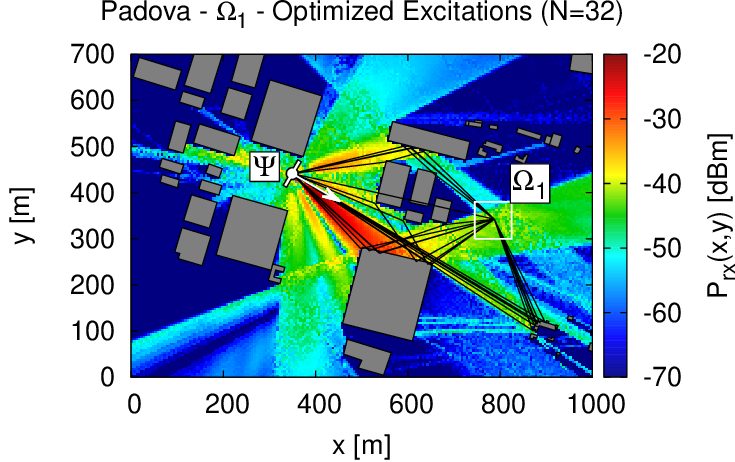}}\tabularnewline
\multicolumn{2}{c}{(\emph{a})}\tabularnewline
\multicolumn{2}{c}{}\tabularnewline
\includegraphics[%
  width=0.40\columnwidth]{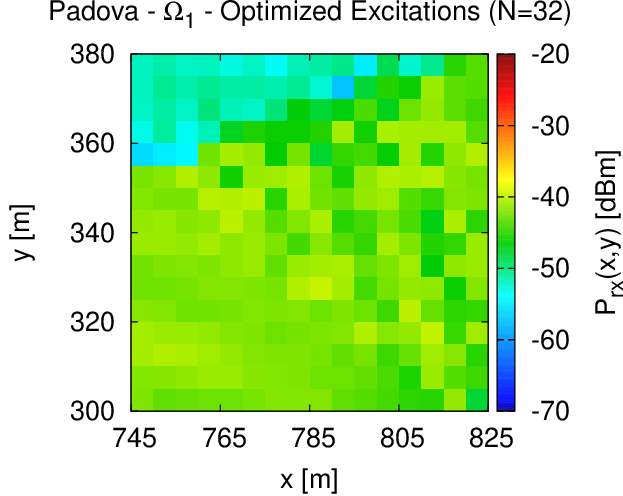}&
\includegraphics[%
  width=0.40\columnwidth]{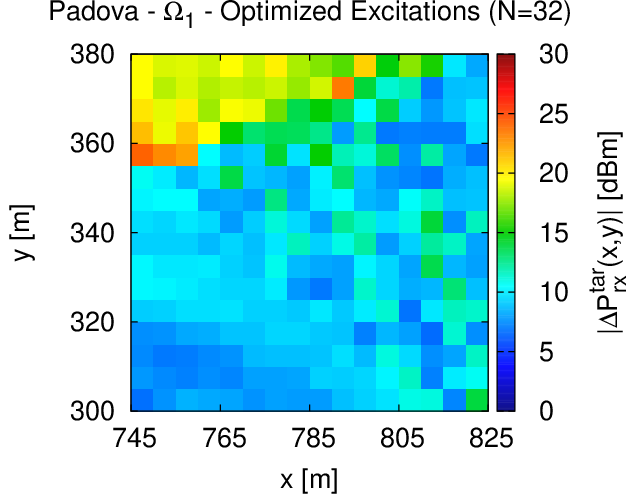} \tabularnewline
(\emph{b})&
(\emph{c})\tabularnewline
&
\tabularnewline
\includegraphics[%
  width=0.40\columnwidth]{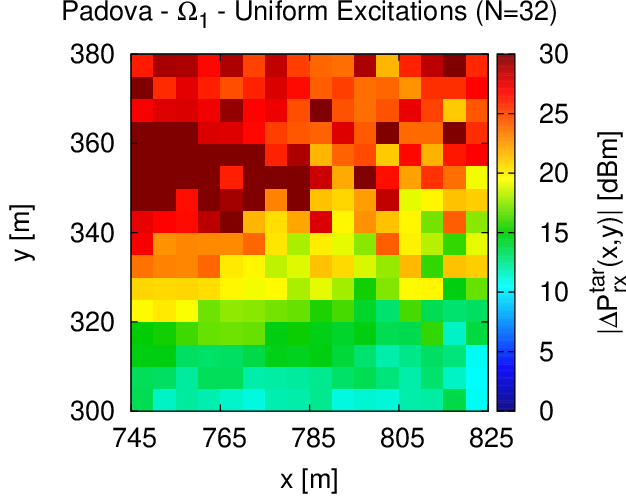} &
\includegraphics[%
  width=0.40\columnwidth]{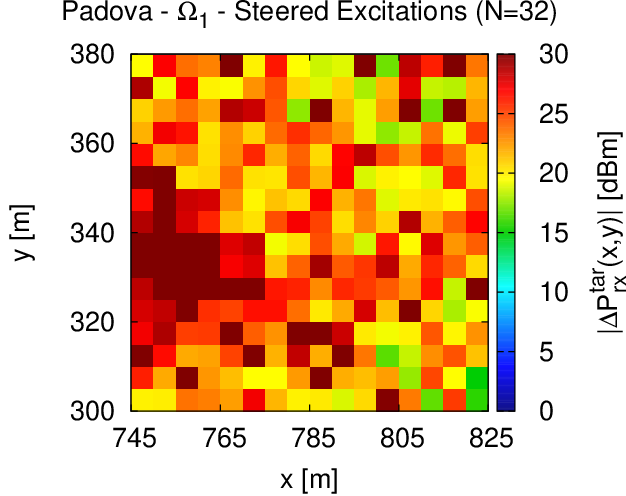} \tabularnewline
(\emph{d})&
(\emph{e})\tabularnewline
\end{tabular}\end{center}

\begin{center}~\vfill\end{center}

\begin{center}\textbf{Fig. 14 - P. Da Ru} \textbf{\emph{et al.}},
{}``An Opportunistic Source Synthesis Method ...''\end{center}

\newpage
\begin{center}~\vfill\end{center}

\begin{center}\begin{tabular}{c}
\includegraphics[%
  width=0.50\columnwidth]{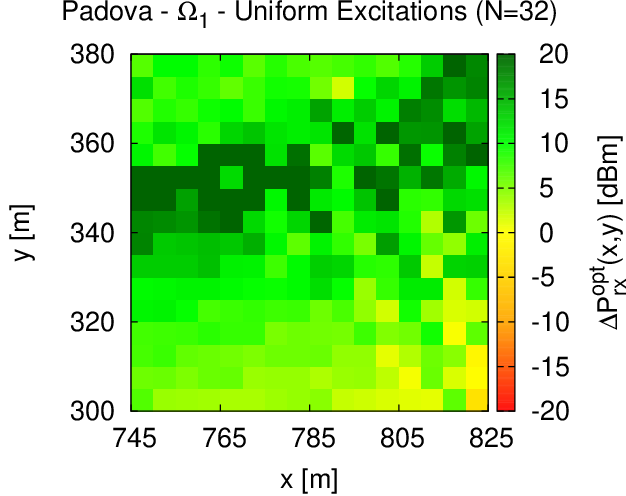}\tabularnewline
(\emph{a})\tabularnewline
\tabularnewline
\includegraphics[%
  width=0.50\columnwidth]{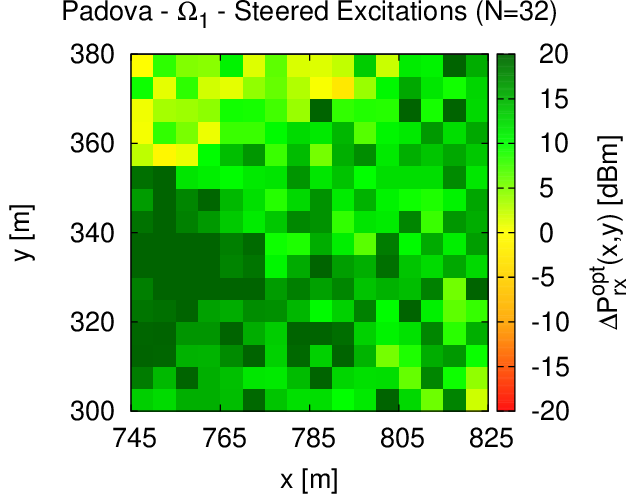}\tabularnewline
(\emph{b})\tabularnewline
\end{tabular}\end{center}

\begin{center}~\vfill\end{center}

\begin{center}\textbf{Fig. 15 - P. Da Ru} \textbf{\emph{et al.}},
{}``An Opportunistic Source Synthesis Method ...''\end{center}

\newpage
\begin{center}~\vfill\end{center}

\begin{center}\includegraphics[%
  width=0.80\columnwidth]{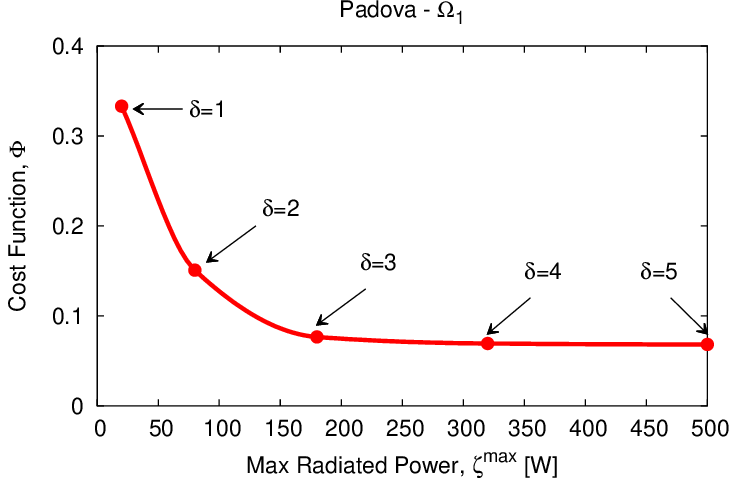}\end{center}

\begin{center}~\vfill\end{center}

\begin{center}\textbf{Fig. 16 - P. Da Ru} \textbf{\emph{et al.}},
{}``An Opportunistic Source Synthesis Method ...''\end{center}

\newpage
\begin{center}~\vfill\end{center}

\begin{center}\begin{tabular}{cc}
\includegraphics[%
  width=0.45\columnwidth]{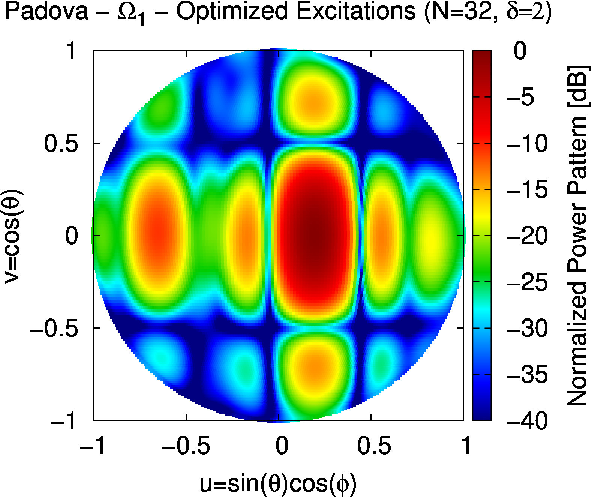}&
\includegraphics[%
  width=0.45\columnwidth]{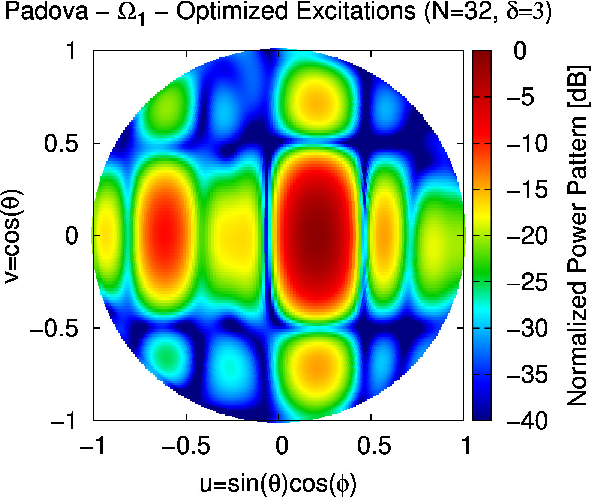}\tabularnewline
(\emph{a})&
(\emph{b})\tabularnewline
&
\tabularnewline
\includegraphics[%
  width=0.45\columnwidth]{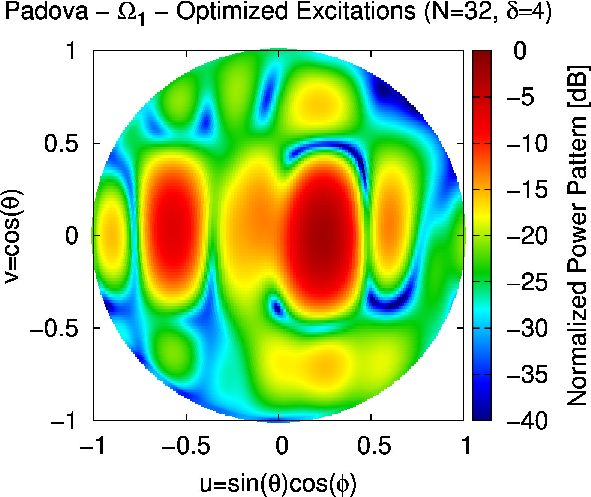}&
\includegraphics[%
  width=0.45\columnwidth]{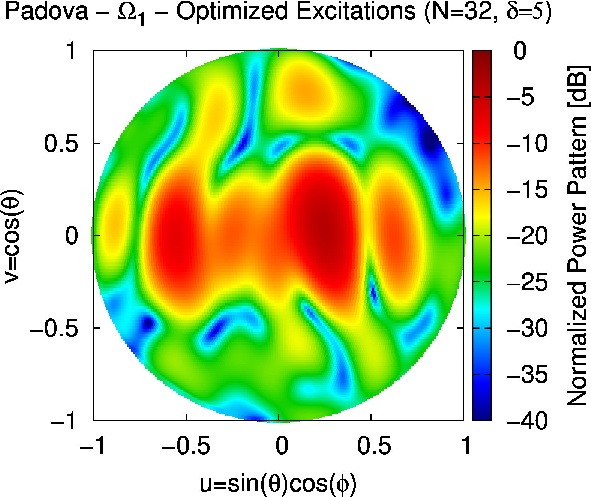}\tabularnewline
(\emph{c})&
(\emph{d})\tabularnewline
\end{tabular}\end{center}

\begin{center}~\vfill\end{center}

\begin{center}\textbf{Fig. 17 - P. Da Ru} \textbf{\emph{et al.}},
{}``An Opportunistic Source Synthesis Method ...''\end{center}

\newpage
\begin{center}~\vfill\end{center}

\begin{center}\begin{tabular}{cc}
\includegraphics[%
  width=0.45\columnwidth]{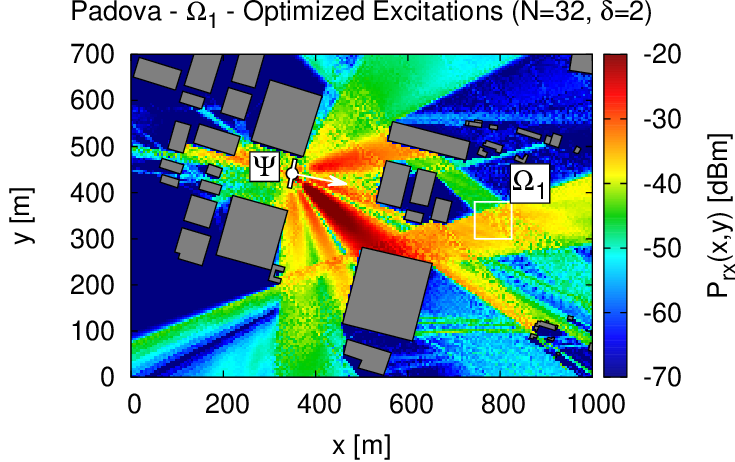}&
\includegraphics[%
  width=0.45\columnwidth]{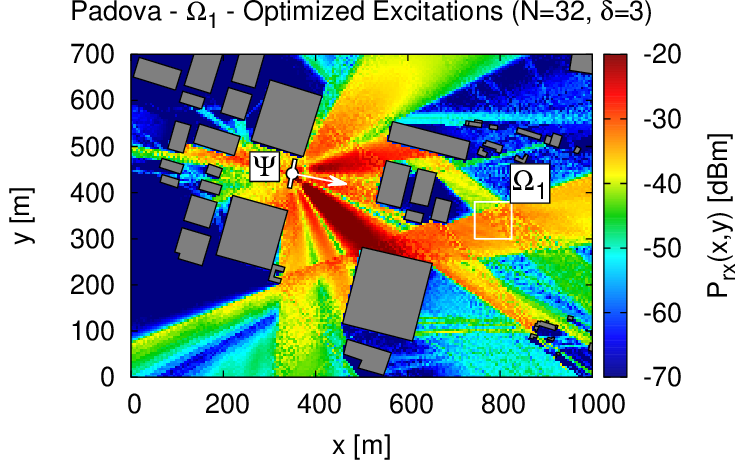}\tabularnewline
(\emph{a})&
(\emph{b})\tabularnewline
&
\tabularnewline
\includegraphics[%
  width=0.45\columnwidth]{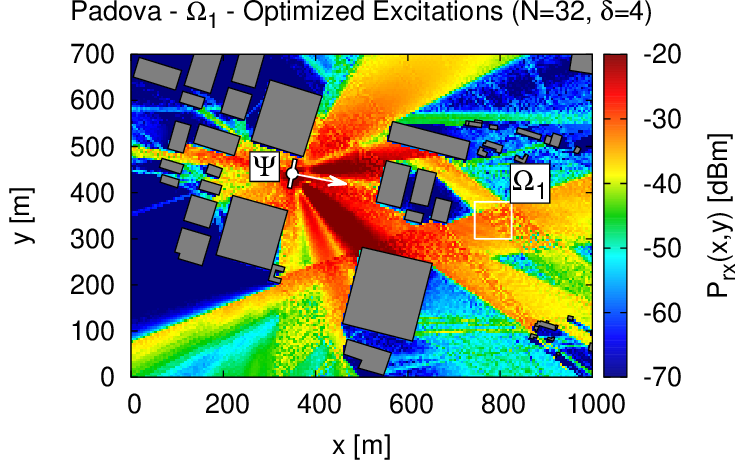}&
\includegraphics[%
  width=0.45\columnwidth]{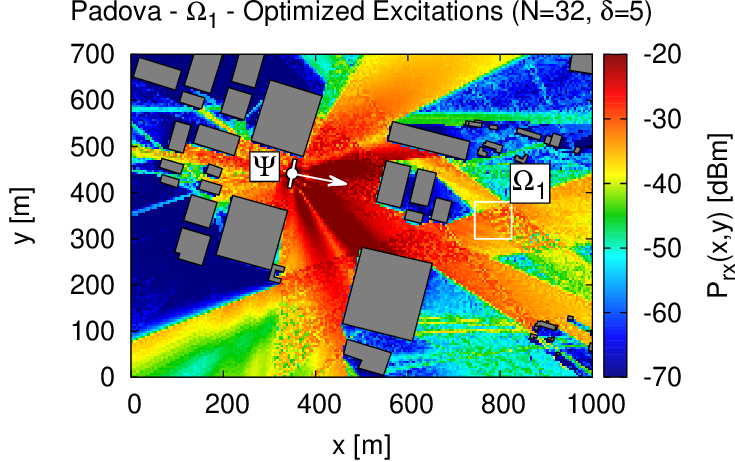}\tabularnewline
(\emph{c})&
(\emph{d})\tabularnewline
\end{tabular}\end{center}

\begin{center}~\vfill\end{center}

\begin{center}\textbf{Fig. 18 - P. Da Ru} \textbf{\emph{et al.}},
{}``An Opportunistic Source Synthesis Method ...''\end{center}

\newpage
\begin{center}~\vfill\end{center}

\begin{center}\begin{tabular}{cc}
\includegraphics[%
  width=0.45\columnwidth,
  height=0.25\columnwidth,
  keepaspectratio]{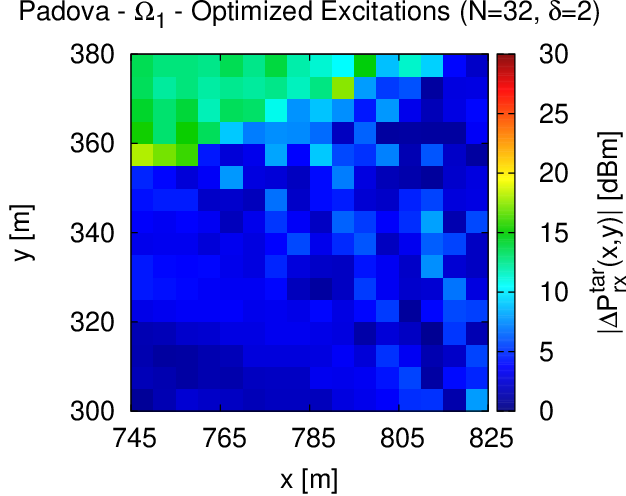} &
\includegraphics[%
  width=0.45\columnwidth,
  height=0.25\columnwidth,
  keepaspectratio]{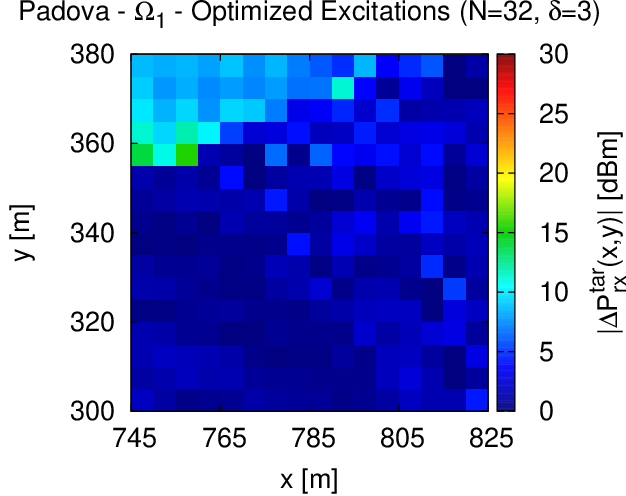} \tabularnewline
(\emph{a})&
(\emph{b})\tabularnewline
&
\tabularnewline
\includegraphics[%
  width=0.45\columnwidth,
  height=0.25\columnwidth,
  keepaspectratio]{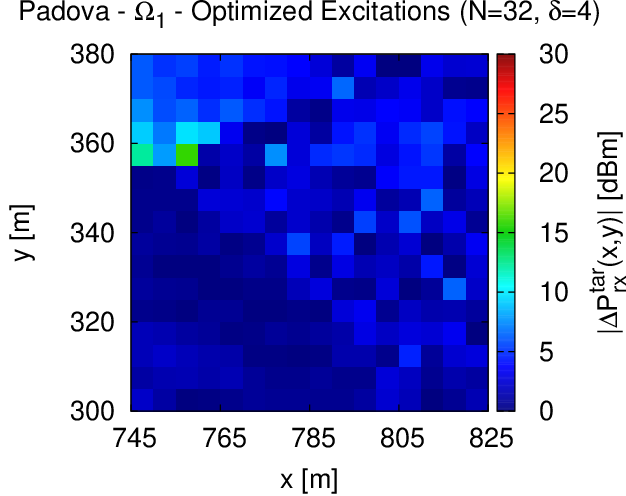} &
\includegraphics[%
  width=0.45\columnwidth,
  height=0.25\columnwidth,
  keepaspectratio]{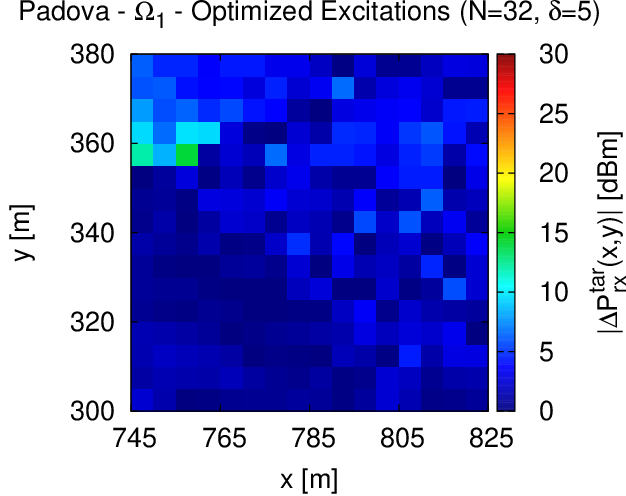} \tabularnewline
(\emph{c})&
(\emph{d})\tabularnewline
\end{tabular}\end{center}

\begin{center}~\vfill\end{center}

\begin{center}\textbf{Fig. 19 - P. Da Ru} \textbf{\emph{et al.}},
{}``An Opportunistic Source Synthesis Method ...''\end{center}

\newpage
\begin{center}~\vfill\end{center}

\begin{center}\begin{tabular}{cc}
\includegraphics[%
  width=0.40\columnwidth]{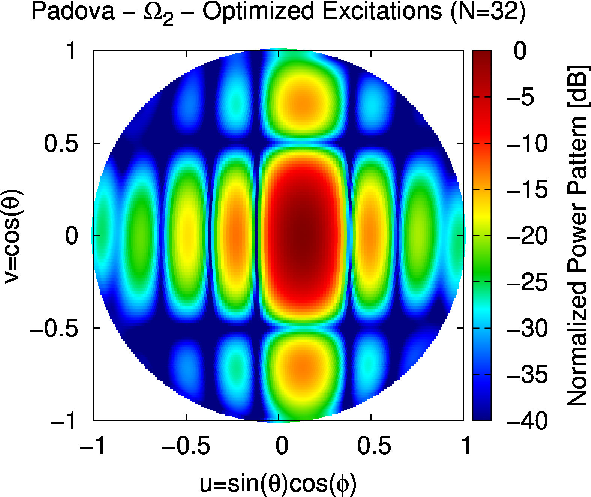}&
\includegraphics[%
  width=0.40\columnwidth]{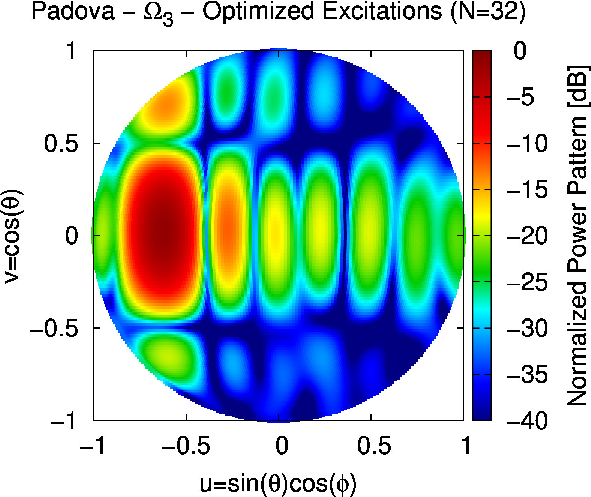}\tabularnewline
(\emph{a})&
(\emph{b})\tabularnewline
&
\tabularnewline
\includegraphics[%
  width=0.48\columnwidth]{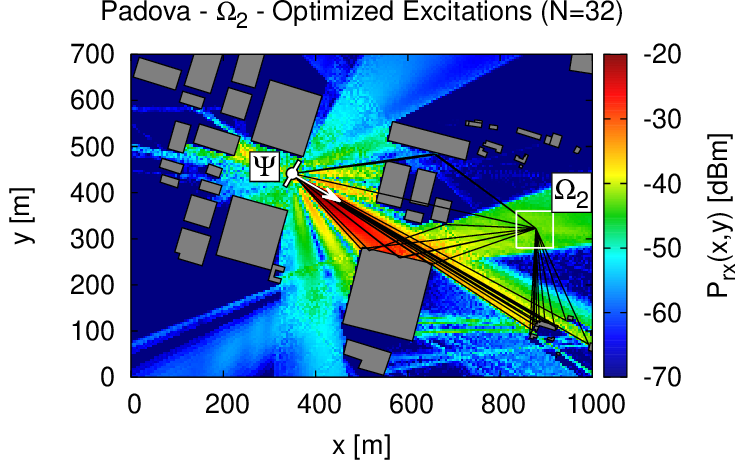}&
\includegraphics[%
  width=0.48\columnwidth]{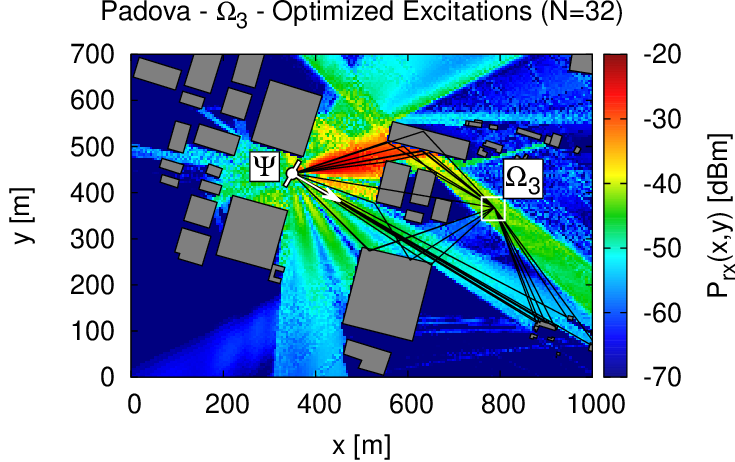}\tabularnewline
(\emph{c})&
(\emph{d})\tabularnewline
&
\tabularnewline
\includegraphics[%
  width=0.35\columnwidth]{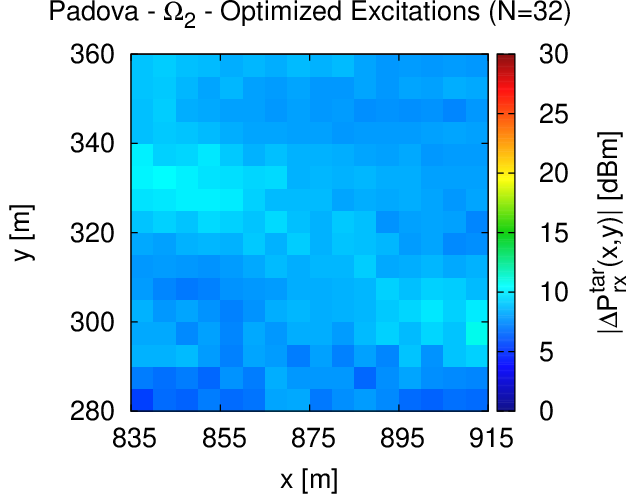} &
\includegraphics[%
  width=0.35\columnwidth]{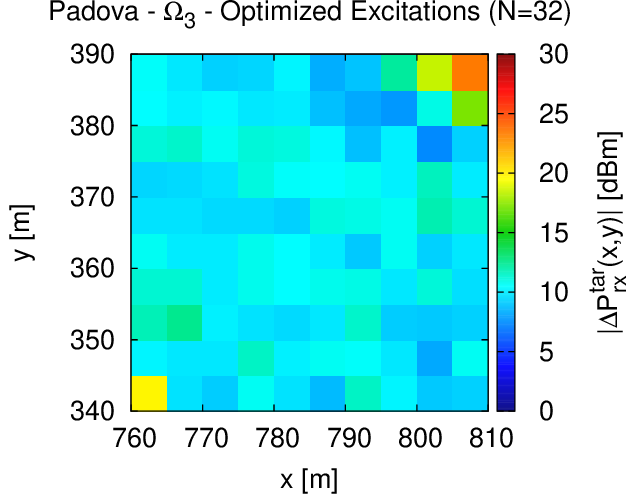} \tabularnewline
(\emph{e})&
(\emph{f})\tabularnewline
\end{tabular}\end{center}

\begin{center}~\vfill\end{center}

\begin{center}\textbf{Fig. 20 - P. Da Ru} \textbf{\emph{et al.}},
{}``An Opportunistic Source Synthesis Method ...''\end{center}

\newpage
\begin{center}~\vfill\end{center}

\begin{center}\begin{tabular}{|c|c|c|c|c|c|}
\hline 
\emph{Test }&
\emph{Statistics}&
\emph{Target}&
$\underline{\beta}=\underline{\beta}^{opt}$&
$\underline{\beta}=\underline{\beta}^{uni}$&
$\underline{\beta}=\underline{\beta}^{ste}$\tabularnewline
\emph{Case}&
($\underline{r}\in\Omega$)&
{[}dBm{]}&
{[}dBm{]}&
{[}dBm{]}&
{[}dBm{]}\tabularnewline
\hline
\hline 
\multicolumn{1}{|c|}{}&
$\min\left\{ P_{rx}\left(\underline{r}\right)\right\} $&
-28.43&
-60.62&
-75.33&
-74.70\tabularnewline
\cline{2-2} \cline{3-3} \cline{4-4} \cline{5-5} \cline{6-6} 
\multicolumn{1}{|c|}{Orgiano}&
$\max\left\{ P_{rx}\left(\underline{r}\right)\right\} $ &
-25.05&
-32.74&
-35.16&
-48.47\tabularnewline
\cline{2-2} \cline{3-3} \cline{4-4} \cline{5-5} \cline{6-6} 
\multicolumn{1}{|c|}{}&
$\mathrm{avg}\left\{ P_{rx}\left(\underline{r}\right)\right\} $&
-26.30&
-38.82&
-41.05&
-54.02\tabularnewline
\hline
\hline 
\multicolumn{1}{|c|}{}&
$\min\left\{ P_{rx}\left(\underline{r}\right)\right\} $&
-35.46&
-57.30&
-75.12&
-69.91\tabularnewline
\cline{2-2} \cline{3-3} \cline{4-4} \cline{5-5} \cline{6-6} 
\cline{2-2} \cline{3-3} \cline{4-4} \cline{5-5} \cline{6-6} 
\multicolumn{1}{|c|}{Padova - $\Omega_{1}$}&
$\max\left\{ P_{rx}\left(\underline{r}\right)\right\} $ &
-31.66&
-39.61&
-44.09&
-49.13\tabularnewline
\cline{2-2} \cline{3-3} \cline{4-4} \cline{5-5} \cline{6-6} 
\cline{2-2} \cline{3-3} \cline{4-4} \cline{5-5} \cline{6-6} 
\multicolumn{1}{|c|}{}&
$\mathrm{avg}\left\{ P_{rx}\left(\underline{r}\right)\right\} $&
-32.98&
-42.91&
-50.89&
-54.79\tabularnewline
\hline
\hline 
\multicolumn{1}{|c|}{}&
$\min\left\{ P_{rx}\left(\underline{r}\right)\right\} $&
-35.98&
-45.68&
-58.72&
-68.05\tabularnewline
\cline{2-2} \cline{3-3} \cline{4-4} \cline{5-5} \cline{6-6} 
Padova - $\Omega_{2}$&
$\max\left\{ P_{rx}\left(\underline{r}\right)\right\} $ &
-33.41&
-41.08&
-40.22&
-47.13\tabularnewline
\cline{2-2} \cline{3-3} \cline{4-4} \cline{5-5} \cline{6-6} 
\multicolumn{1}{|c|}{}&
$\mathrm{avg}\left\{ P_{rx}\left(\underline{r}\right)\right\} $&
-34.42&
-42.70&
-46.10&
-53.58\tabularnewline
\hline
\hline 
&
$\min\left\{ P_{rx}\left(\underline{r}\right)\right\} $&
-33.31&
-56.39&
-75.12&
-70.65\tabularnewline
\cline{2-2} \cline{3-3} \cline{4-4} \cline{5-5} \cline{6-6} 
\multicolumn{1}{|c|}{Padova - $\Omega_{3}$}&
$\max\left\{ P_{rx}\left(\underline{r}\right)\right\} $ &
-31.87&
-40.16&
-51.95&
-50.43\tabularnewline
\cline{2-2} \cline{3-3} \cline{4-4} \cline{5-5} \cline{6-6} 
\multicolumn{1}{|c|}{}&
$\mathrm{avg}\left\{ P_{rx}\left(\underline{r}\right)\right\} $&
-32.54&
-42.76&
-57.97&
-55.75\tabularnewline
\hline
\end{tabular}\end{center}

\begin{center}~\vfill\end{center}

\begin{center}\textbf{Tab. I - P. Da Ru} \textbf{\emph{et al.}}, {}``An
Opportunistic Source Synthesis Method ...''\end{center}

\newpage
\begin{center}~\vfill\end{center}

\begin{center}\begin{tabular}{|c|c|c|c|c|}
\hline 
\emph{Statistics }&
$\delta=2$&
$\delta=3$&
$\delta=4$&
$\delta=5$\tabularnewline
($\underline{r}\in\Omega_{1}$)&
{[}dBm{]}&
{[}dBm{]}&
{[}dBm{]}&
{[}dBm{]}\tabularnewline
\hline
\hline 
$\min\left\{ P_{rx}^{opt}\left(\underline{r}\right)\right\} $&
-51.31&
-47.47&
-47.76&
-45.96\tabularnewline
\hline 
$\max\left\{ P_{rx}^{opt}\left(\underline{r}\right)\right\} $&
-33.60&
-30.31&
-28.11&
-27.59\tabularnewline
\hline 
$\mathrm{avg}\left\{ P_{rx}^{opt}\left(\underline{r}\right)\right\} $&
-36.88&
-33.68&
-32.84&
-32.71\tabularnewline
\hline
\end{tabular}\end{center}

\begin{center}~\vfill\end{center}

\begin{center}\textbf{Tab. II - P. Da Ru} \textbf{\emph{et al.}},
{}``An Opportunistic Source Synthesis Method ...''\end{center}
\end{document}